\documentclass[prx,aps,showpacs,preprintnumbers,
amsmath,amssymb,superscriptaddress,longbibliography,colorlinks=true, a4paper=true, pdfstartview=FitV,linkcolor=blue, citecolor=blue, urlcolor=blue,twocolumn]{revtex4-1}

\usepackage[T1]{fontenc}		
\usepackage[english]{babel}		
\usepackage[utf8]{inputenc}	
\usepackage{times}
\usepackage{tikz}
\usepackage{tensor}
\usetikzlibrary{knots}

\usepackage{amsmath}			
\usepackage{amssymb}			
\usepackage{bbm}				

\usepackage{hyperref}
\usepackage{pbox}

\usepackage{graphicx}			
\usepackage{graphics}			
\DeclareGraphicsExtensions{.pdf}
\usepackage{color}	
\usepackage{makecell}

\newcommand{\bea}{\begin{eqnarray}}
\newcommand{\eea}{\end{eqnarray}}

\newcommand{\bsigma} {\boldsymbol{\sigma}}

\DeclareMathOperator{\HH}{\mathcal{H}}

\DeclareMathOperator{\sgn}{sgn}

\allowdisplaybreaks

\addto\captionsenglish{}

\begin{document}
\title{
Magneto-optical Hall response in generic Weyl semimetals}

\author{Marcus St{\aa}lhammar}
\email{marcus.stalhammar@su.se}
\affiliation{Nordita, KTH Royal Institute of Technology and Stockholm University, Hannes Alfv\'ens v\"ag 12, SE-106 91 Stockholm, Sweden}

\begin{abstract}
Weyl semimetals are predicted to host signature magneto-optical properties sourced by their peculiar Landau level structure, including the chiral level.
Analytical studies are often leaving out the Hall component of the conductivity due to its complicated nature, and even though the chiral anomaly requires Weyl nodes to come in charge-conjugate pairs, toy-models hosting only one node are considered almost exclusively; numerical studies including several Weyl nodes are on the other hand often limited to high-field quantum limits or DC studies.
Here, I present a twofold purpose study, where I a) analytically derive a closed-form expression also for the Hall conductivity of a generic Weyl semimetal using linear response theory, and b) apply this general framework to evaluate the transverse conductivity components for Weyl systems with two nodes. I study how various model parameters, including the tilt, momentum separation, and energy location of the nodes, as well as the chemical potential affect the magneto-optical conductivity, and complement these studies with deriving an analytical expression for the DC Hall conductivity, which is also evaluated in various systems. Including a chiral pair of nodes result two important differences compared to earlier studies; the contribution from the chiral level is equal in size but opposite at the two nodes, making the net contribution to disappear; the energy scales at which intraband transitions occur is smeared out and approaches that of interband transitions, strengthening the hypothesis that intraband transitions mask signature optical features in materials. 
This general formalism can be applied for a large family of generic Weyl semimetals, and comprise an important piece towards unravelling the source of the mismatch between theoretical predictions and experimental observations in candidate materials.
\end{abstract}
\maketitle


\section{Introduction} \label{sec:introduction}
Topological phases of matter have been extensively studied both from a theoretical and experimental perspective since the observation of the quantum Hall effect in the 1980s \cite{Klitzing1980}. 
This has resulted in the tenfold way classification \cite{Altland1997}, a thorough framework for understanding of gapped topological insulators \cite{hasankane} and superconductors \cite{qizhang} as well as gapless systems as the semimetallic state of graphene in two dimensions \cite{goerbig}, and Weyl semimetals in three dimensions \cite{weylreview}. 
Apart from proving fruitful within condensed matter physics, these studies have profound consequences in other areas of physics, where one paramount example is the realization of Weyl fermions as quasiparticles excitations in the vicinity of the point-like band intersections in Weyl semimetals \cite{XBAN2015,LWFM2015,LWYRFJS2015}; these elusive particles were originally theoretically predicted by Hermann Weyl in 1929 in the context of particle physics \cite{W1929}, and are yet to be observed as actual fundamental particles. 
The intersections, naturally known as Weyl points or Weyl nodes, appear in two different types distinguished by the behaviour of the dispersion in the vicinity of the degeneracy \cite{weylreview}. 
In type I systems the valence and conduction bands merely touch the Fermi level, forming a proper semimetallic state with a point-like Fermi surface \cite{WTVS2011,HQ2013,HZLW2015}, while type II nodes come with a finite Fermi surface as a consequence of the dispersion being overtilted \cite{BLTMU2015,SGWWTDB2015}. 

The existence of Weyl nodes have furthermore indicated that Weyl semimetals ought to host unique transport properties related to electromagnetic response. 
Salient examples are negative magnetoresistence \cite{AXFT2014,KSXMY2017,Zetal2016} and the chiral anomaly \cite{ZXBY2016,NM1983,XPYQ2015,SY2012,G2012,ZB2012,GT2013,PGAPV2014,KGM2015,KGM2017,BKS2018,Huangetal2015}, but do also include signature magneto-optical properties. 
The magneto-optical conductivity, for example, is predicted to host a series of asymmetrically spaced peaks, caused by optical transitions between Landau levels \cite{AC2013,Tchoumakov2016,Yadav2022,Stalhammar20}, something that initially was believed to work as an experimental signature identifying Weyl semimetals \cite{weylreview}. 
Nowadays it is known that theoretical predictions are at odds with experimental observations \cite{PD2020,XZC2018,PGW2020}, something that has been suggested is due to the existence of trivial Fermi pockets, i.e., regions where the bands cross the Fermi level not in direct connection to a Weyl node \cite{Stalhammar20}.

Studies on magneto-optical transport properties in Weyl semimetals are usually carried out with two rather different approaches. 
Either, simplified pictures including systems with only one Weyl node are studied analytically \cite{AC2013,Tchoumakov2016,Yadav2022,Stalhammar20,TBK2018,TBUK2017,YYY2016,UB2016}, or the studies rely exclusively on numerical simulations of lattice systems, see Ref.~\cite{Xiong2022} for a recent study. Additional examples include conductivity in Weyl semimetal sheets \cite{Chang2023}, high-field quantum limits \cite{Lu2015} and magneto-optical transport in the DC-limit \cite{Li2016,Wang2017,Li2020}. 
All these works have provided fruitful insights regarding transport phenomena and the quantum hall effect in three dimensions, but in one way or the other lack either a realistic or generic setup. 
Therefore, we in this work investigate the effects of having systems with a chiral pair of Weyl nodes, including fourth-order momentum terms in the dispersion. 
We investigate how the interplay between the nodes, such as its tilts and position in energy and momentum space, affect the transport properties, and also how these are connected to the appearance of trivial and non-trivial Fermi pockets. 
We go further beyond previous studies by expanding the general analytical framework developed in Ref.~\cite{Stalhammar20}, where a closed-form expression for the dissipative components of the conductivity tensor are derived, to also include the reactive components. 
This allows us to not only study transport in terms of the real part of $\sigma_{xx}$, but also the Hall conductivity, the real part of $\sigma_{xy}$, at any frequencies. 
Through the same technique, we also analytically derive the DC Hall conductivity, and we display how these analytical expressions serve as a tool for predicting physical features directly from the band structure.

The study shows that a charge-conjugate pair of Weyl nodes makes the contribution from the chiral level to cancel, and further indicate that the energy scales at which intraband transitions appear are smeared out, strengthening the hypothesis that these are responsible for the disagreements between theory and experiments.
This analytical framework thus comprise a natural step towards an agreement between theory and experiment, and can in principle work as a tool to make transport predictions for material candidates as long as the corresponding band structure is provided (e.g., using DFT). 
It furthermore contributes to an increased and more complete theoretical understanding of transport features in three dimensions, including the three dimensional quantum Hall effect, thanks to the corresponding analytical expressions for the respective quantity.

The rest of the article is structured as follows. 
We set the stage in Sec.~\ref{sec:genbackground} and provide a brief background of previous works and analytical techniques. 
In particular, we show how the Landau level splitting works in generic Weyl systems, and provide a simplification of the expressions for the dissipative conductivity components, before we turn to deriving the closed-form expressions for the corresponding reactive components and the DC Hall conductivity. 
In Sec.~\ref{sec:results}, the formalism is applied to various Weyl systems with two Weyl nodes. 
We investigate how the tilt of the respective nodes, and hence the presence of trivial and non-trivial Fermi pockets, is manifest in transport. In particular, we study the different features present in the type-I and type-II regimes, try to identify the interplay between these two regimes. 
We discuss the results and their relevance in Sec.~\ref{sec:discussion}, and provide additional plots for when the two Weyl nodes are at different energies, and for the corresponding DC-conductivity. 
We summarise and conclude in Sec.~\ref{sec:summary}.


\section{Magneto-optical conductivity for generic Weyl semimetals} \label{sec:genbackground}
Take as starting point a generalized Weyl Hamiltonian,
\begin{equation}
    \HH = \hbar v_F \left[ h(k_z)\sigma^0+ k_x\sigma^x+k_y\sigma^y +g(k_z) \sigma^z\right],
\end{equation}
where $\mathbf{k}=(k_x,k_y,k_z)$ is the lattice momentum, $\bsigma = (\sigma^x,\sigma^y,\sigma^z)$ the Pauli  matrices, $\sigma^0$ the $2\times 2$ identity matrix, $v_F$ the Fermi velocity and $h(k_z)$ and $g(k_z)$ continuously differentiable functions of $k_z$ only. 
The function $h(k_z)$ introduces a tilt in the $k_z$-direction, and $g(k_z)$ specifies the position of the Weyl nodes in the spectrum.

To study magneto-optical transport, we couple the Hamiltonian to an external magnetic field through minimal coupling. 
For concreteness, the magnetic field is aligned with the direction of the tilt, i.e., $\mathbf{B}=(0,0,B)$, which in the Landau gauge corresponds to introducing a vector potential on the form $\mathbf{A} = (0,Bx,0)$. 
The minimal coupling transforms the momentum as $\hbar k_i\to \Pi_i = \hbar k_i - \frac{e}{c} A_i$. 
By introducing creation and annihilation operators as $a = \frac{l_B}{\sqrt{2}\hbar}\left(\Pi_x-i\Pi_y\right)$ and $a^{\dagger}=\frac{l_B}{\sqrt{2}\hbar} \left(\Pi_x+i\Pi_y\right)$, which satisfy $\left[a,a^{\dagger}\right]=1$, the system under the influence of a magnetic field can be effectively described by,
\begin{equation}
    H = \hbar v_F \begin{pmatrix} h(k_z)+g(k_z) & \frac{\sqrt{2}}{l_B}a^{\dagger} \\\frac{\sqrt{2}}{l_B}a & h(k_z)-g(k_z) \end{pmatrix}.
\end{equation}
The corresponding eigenvalues are
\begin{align}
    E_{n,\lambda}(k_z,l_B) &= \hbar v_F \left[h(k_z)+\lambda \sqrt{g^2(k_z)+\frac{2n}{l^2_B}}\right], \quad n\geq 1, \label{eq:LLn}
    \\
    E_0(k_z) &= \hbar v_F \left[h(k_z)+g(k_z)\right] \label{eq:LLc},
\end{align}
with $\lambda=\pm 1$, and the eigenstates cast the form $\psi_{n,\lambda}(k_z,l_B) = \begin{pmatrix}
    \lambda u_{n,\lambda}(k_z,l_B)\\ v_{n,\lambda}(k_z,l_B)
\end{pmatrix}$, with 
\begin{align}
    u_{n,\lambda}(k_z,\l_B) &= \sqrt{\frac{1}{2} \left[1+\frac{g(k_z)}{\lambda\sqrt{g^2(k_z)+\frac{2n}{l^2_B}}}\right]}, \quad n\geq 1,
    \\
    v_{n,\lambda}(k_z,\l_B) &= \sqrt{\frac{1}{2} \left[1-\frac{g(k_z)}{\sqrt{g^2(k_z)+\frac{2n}{l^2_B}}}\right]}, \quad n\geq 1,
    \\
    \psi_{0} &= \begin{pmatrix}1\\0\end{pmatrix}.
\end{align}

\subsection{Dissipative components of the response function} \label{sec:disscomp}
The magneto-optical conductivity is, in the one-loop approximation, calculated using linear response theory and specifically expressed in terms of the response function (in what follows, we will omit certain dependencies of parameters, and re-introduce them when appropriate),
\begin{align} \label{eq:linres}
    &\chi_{\mu \nu}(\omega) = \nonumber
    \\
    & \frac{1}{2\pi l^2_B}\sum_{n,n'}\sum_{\lambda,\lambda'}\int_{-\infty}^{+\infty} \frac{dk_z}{2\pi}\frac{f\left[E_{n,\lambda}(k_z)\right]-f\left[E_{n',\lambda'}(k_z)\right]}{\hbar \omega + E_{n,\lambda}(k_z)-E_{n',\lambda'}(k_z)+i\epsilon} \nonumber
    \\
    &\times \langle \psi_{n,\lambda}(k_z)|j_{\mu}|\psi_{n',\lambda'}(k_z)\rangle\langle\psi_{n',\lambda'}(k_z)|j_{\nu}|\psi_{n,\lambda}(k_z)\rangle,
\end{align}
with $f(E) = \frac{1}{1+e^{\beta\left(E-\mu\right)}}$ the Fermi-Dirac distribution, $\beta^{-1}=k_B T$, $T$ the temperature and $k_B$ the Boltzmann constant. Small, but finite, and positive $\epsilon$ regulates the integral, and can be thought of as an inverse impurity scattering lifetime $\tau$ via $\epsilon=\frac{\hbar}{2\tau}$. The current operators $j_{\mu}$ are given by
\begin{equation}
    j_{\mu} = \frac{e}{\hbar} \frac{\partial H}{\partial \Pi_{\mu}}.
\end{equation}

In Ref.~\cite{Stalhammar20}, analytical expressions for the dissipative parts of $\chi_{xx}(\omega)$ and $\chi_{xy}(\omega)$ for generic Weyl systems were derived. 
We do not repeat this derivation here, but instead further simplify these expressions, with the final goal of also deriving a closed-form analytical expression for the reactive parts of the response function. 
The final result of Ref.~\cite{Stalhammar20} can, after some straight-forward algebra, be put on the form
\begin{align}
    &\text{Im}\left[\chi_{xx}(\omega)\right]=-\frac{e^2v_F^2}{16\pi l^2_B}\sum_{n=0}^{n_{\text{max}}}\sum_{i=1}^{2m} \nonumber
    \\
    &\left\{ A_n^{++}(k_i)\left[\frac{\frac{2n+1}{l^2_B}-\frac{\omega^2}{2v_F^2}}{\left|\tilde{g}(k_i)g'(k_i)\right|}\right]\theta\left(\frac{2 v_F^2}{l_B^2}-\omega^2\right)\sgn{(-\omega)}\right.\nonumber
    \\
    &+A^{--}_n(k_i)\left[\frac{\frac{2n+1}{l^2_B}-\frac{\omega^2}{2v_F^2}}{\left|\tilde{g}(k_i)g'(k_i)\right|}\right]\theta\left(\frac{2 v_F^2}{l_B^2}-\omega^2\right)\sgn{(\omega)}\nonumber
    \\
    & +A_n^{+-}(k_i)\left[\frac{\frac{\omega^2}{2v_F^2}-\frac{2n+1}{l^2_B}}{\left|\tilde{g}(k_i)g'(k_i)\right|}\right]\theta\left(\omega^2-\frac{2 v_F^2}{l_B^2}\right)\sgn{(\omega)}\nonumber
    \\
    &\left. +A^{-+}_n(k_i)\left[\frac{\frac{\omega^2}{2v_F^2}-\frac{2n+1}{l^2_B}}{\left|\tilde{g}(k_i)g'(k_i)\right|}\right]\theta\left(\omega^2-\frac{2 v_F^2}{l_B^2}\right)\sgn{(-\omega)} \right\},
    \end{align}
    \begin{align}
    &\text{Re}\left[\chi_{xy}(\omega)\right]=-\frac{e^2v_F^2}{16\pi l^2_B}\sum_{n=0}^{n_{\text{max}}}\sum_{i=1}^{2m} \nonumber
    \\
    &\left\{ A_n^{++}(k_i)\left[\frac{\frac{2n+1}{l^2_B}-\frac{\omega^2}{2v_F^2}}{\left|\tilde{g}(k_i)g'(k_i)\right|}\right]\theta\left(\frac{2 v_F^2}{l_B^2}-\omega^2\right)\right.\nonumber
    \\
    &+A^{--}_n(k_i)\left[\frac{\frac{2n+1}{l^2_B}-\frac{\omega^2}{2v_F^2}}{\left|\tilde{g}(k_i)g'(k_i)\right|}\right]\theta\left(\frac{2 v_F^2}{l_B^2}-\omega^2\right) \nonumber
    \\
    & +A_n^{+-}(k_i)\left[\frac{\frac{\omega^2}{2v_F^2}-\frac{2n+1}{l^2_B}}{\left|\tilde{g}(k_i)g'(k_i)\right|}\right]\theta\left(\omega^2-\frac{2 v_F^2}{l_B^2}\right)\nonumber
    \\
    &\left. +A^{-+}_n(k_i)\left[\frac{\frac{\omega^2}{2v_F^2}-\frac{2n+1}{l^2_B}}{\left|\tilde{g}(k_i)g'(k_i)\right|}\right]\theta\left(\omega^2-\frac{2 v_F^2}{l_B^2}\right)\right\},
\end{align}
with
\begin{align}
    g(k_i) &= \pm\sqrt{\frac{v_F^2}{\omega^2l_B^4}-\frac{2n+1}{l^2_B}+\frac{\omega^2}{4v_F^2}}, \label{eq:deltak}
    \\
    \tilde{g}(k_i) &= \left|\frac{\omega}{v_F}\right| g(k_i), 
    \\
    A_{n}^{\lambda\lambda'}(k_i) &= f\left[E_{n,\lambda}(k_i)\right]-f\left[E_{n+1,\lambda'}(k_i)\right],
    \\
    n_{\text{max}} &
    = \left\lfloor\frac{\left(2 v_F^2-\omega^2l^2_B\right)^2}{8 v_F^2\omega^2l_B^2}\right\rfloor,
\end{align}
$\theta(x)$ the Heaviside step function, and $k_i$ the value of $k_z$ that satisfies
Eq.~\eqref{eq:deltak}. We devote Appendix~\ref{appendix} to the calculational details. Here it is important to keep in mind that $g(k_i)$ and $k_i$, and hence also $g'(k_i)$, are bound to be real, otherwise they do not represent the physical problem that we are considering; they are forced to be real since $g(k_z)$ appear in the energy eigenvalues in Eqs.~\eqref{eq:LLn} and \eqref{eq:LLc}. This will be important later on.
Further note that $h(k_z)$ enters through the factors $A_n^{\lambda\lambda'}$, since their definition includes the dispersion relation.

The Kramers-Kronig relations tells us that the reactive and dissipative components of the response functions are given in terms of each other as
\begin{align}
    \text{Re}\left[\chi_{\mu\nu}(\omega)\right] &= \text{PV}\int_{-\infty}^{+\infty}\frac{d\omega'}{\pi} \frac{\text{Im}\left[\chi_{\mu\nu}(\omega')\right]}{\omega'-\omega} \label{eq:KrKr1},
    \\
    \text{Im}\left[\chi_{\mu\nu}(\omega)\right] &= -\text{PV}\int_{-\infty}^{+\infty}\frac{d\omega'}{\pi} \frac{\text{Re}\left[\chi_{\mu\nu}(\omega')\right]}{\omega'-\omega},\label{eq:KrKr2}
\end{align}
where PV denotes the Cauchy principal value of the corresponding integral. 
This means that to derive analytical expressions for $\text{Re}\left[\chi_{xx}(\omega)\right]$ and $\text{Im}\left[\chi_{xy}(\omega)\right]$, respectively, we need to evaluate the corresponding principal value integrals. 
This will be topic of the following subsection. 
Finally, the conductivity tensor is given by the Kubo formula,
\begin{equation}
    \sigma_{\mu\nu}(\omega) = \frac{1}{i\omega}\left[\chi_{\mu\nu}(\omega)-\chi_{\mu\nu}(0)\right].
\end{equation}

For generic Weyl systems, the Landau levels are completely non-degenerate, meaning that $E_{n,\lambda}(k_z)\neq E_{n',\lambda'}(k_z)$ for every $n\neq n'$, every $\lambda$ and $\lambda'$, and all $k_z$. 
Therefore, $\chi_{\mu\nu}(0)$ will vanish in the clean limit, $\epsilon\to 0$, and for $\omega\neq 0$, the conductivity components for clean Weyl systems can be put on the form
\begin{align}
    \sigma_{xx}(\omega) &= \frac{1}{\omega}\left\{\text{Im}\left[\chi_{xx}(\omega)\right] - i \text{Re}\left[\chi_{xx}(\omega)\right]\right\} \nonumber
    \\
    &= \frac{1}{\omega}\left\{\text{Im}\left[\chi_{xx}(\omega)\right]-i\text{PV}\int_{-\infty}^{+\infty}\frac{d\omega'}{\pi} \frac{\text{Im}\left[\chi_{\mu\nu}(\omega')\right]}{\omega'-\omega}\right\},
    \\
    \sigma_{xy}(\omega) &= \frac{1}{\omega}\left\{\text{Im}\left[\chi_{xy}(\omega)\right]-i\text{Re}\left[\chi_{xy}(\omega)\right]\right\} \nonumber
    \\
    &=-\frac{1}{\omega}\left\{ \text{PV}\int_{-\infty}^{+\infty}\frac{d\omega'}{\pi} \frac{\text{Re}\left[\chi_{\mu\nu}(\omega')\right]}{\omega'-\omega} + i\text{Re}\left[\chi_{xy}(\omega)\right]\right\}.
\end{align}
This makes it clear that to obtain the Hall conductivity, the principal value has to be computed, which is the main motivation for finding a closed-form expression for it.

\subsection{Reactive components of the response function} \label{sec:reactcomp}
\begin{figure*}[hbt!]
\centering

\includegraphics[width=\textwidth]{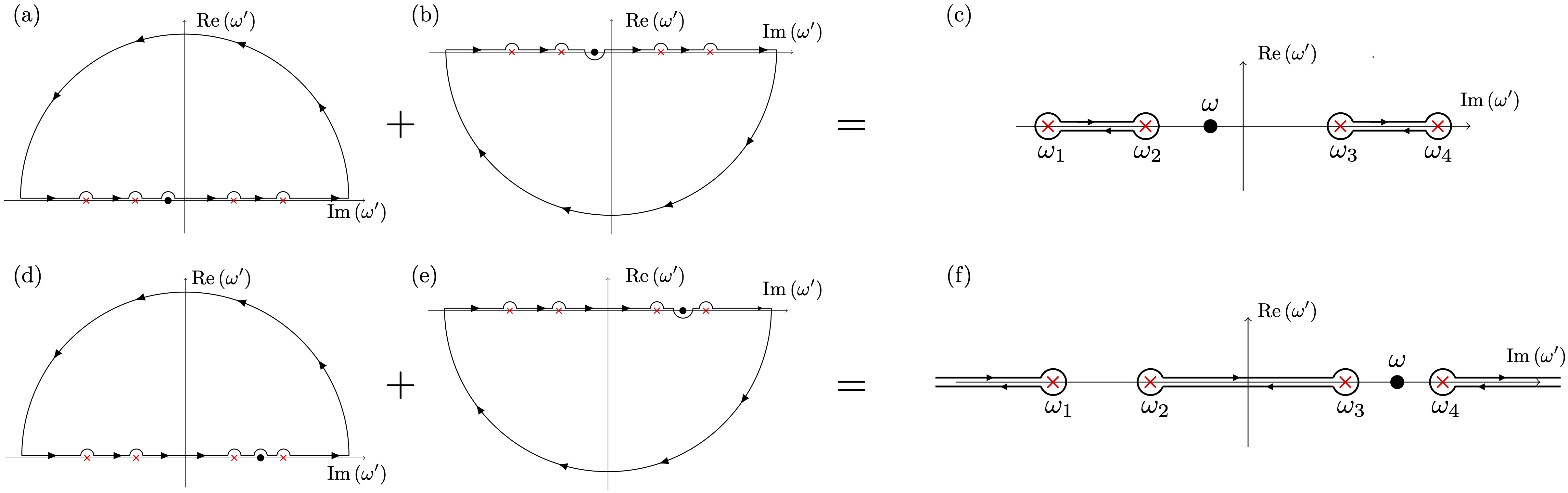}

\caption{Integration contours that schematically shows how to evaluate the principal values in Eqs.~\eqref{eq:PVxx} and \eqref{eq:PVxy}, depending on the position of the pole $\omega'=\omega$. When the point $\omega'=\omega$ has an even number of poles to its left, a series of finite branch cuts are made between neighbouring branch points $\omega_j$, and the sum of the integration contours and panel (a) and (b) becomes a series of dogbone contours around the corresponding branch cuts, as depicted in panel (c). When $\omega$ instead has an odd number of poles to its left, the sum of the contours in (d) and (e) deforms into two hairpin contours between $-\infty$ and the left-most branch point, and the right-most branch point to $+\infty$, denoted $\omega_1$ and $\omega_4$ in (f), respectively, and a sum of finite dogbone contours between rest of neighbouring branch points, as depicted in panel (f). 
 \label{fig:integration contours} }
\end{figure*}

The reactive components of the response functions are directly related to the dissipative components through the Kramers-Kronig relations, Eqs.~\eqref{eq:KrKr1} and \eqref{eq:KrKr2}. To evaluate the corresponding principal values, we first consider the zero-temperature limit, where the factors $A_{n}^{\lambda\lambda'}(k_i)$ are
\begin{equation}
    A_n^{\lambda\lambda'}(k_i) \xrightarrow[]{T\to 0} \theta\left[\mu-E_{n,\lambda}(k_i)\right]-\theta \left[\mu-E_{n+1,\lambda'}(k_i)\right].
\end{equation}
In this limit, the dissipative components of the response functions will host singularities only when $\omega'=\omega$‚ when $g\left[k_i(\omega',n)\right]=0$, or when $g'\left[k_i(\omega',n)\right]=0$. 
The singularity at $\omega'=\omega$ is a simple pole located at the real axis, while the remaining singularities are square-root branch points, the location of which we need to identify. 
For this, we first recall that for causality reasons, the response function is analytic in the upper half complex $\omega$-plane, meaning that for $\text{Im}\left(\omega'\right)>0$, $\chi_{\mu\nu}(\omega')$ is analytic. 
Furthermore, the branch points related to $g\left[k_i(\omega',n)\right]=0$ and $g'\left[k_i(\omega',n)\right]=0$ necessarily come in respective complex conjugate pairs, since both $g(k_i)$ and $g'(k_i)$ are purely real [recall that $g(k_z)$ defines the energy eigenvalues, which are bound to be real]. But since there are no poles in the upper half complex $\omega'$ plane, $g=0$ and $g'=0$ cannot be satisfied for any complex $\omega'$. This means that all singularities, both poles and branch cuts, are constrained to the real line. Using this along with the relations
\begin{align}
    &\quad \text{PV} \int_{-\infty}^{+\infty}\frac{d\omega'}{\pi}\frac{\text{Re}\left[\chi_{\mu\nu}(\omega')\right]}{\omega'-\omega} \nonumber
    \\
    &= \frac{1}{2} \int_{-\infty}^{+\infty}\frac{d\omega'}{\pi}\left\{\frac{\text{Re}\left[\chi_{\mu\nu}(\omega')\right]}{\omega'-\omega-i\epsilon}+\frac{\text{Re}\left[\chi_{\mu\nu}(\omega')\right]}{\omega'-\omega+i\epsilon}\right\} \label{eq:KrKrIm},
    \\
    &\quad \text{PV} \int_{-\infty}^{+\infty}\frac{d\omega'}{\pi}\frac{\text{Im}\left[\chi_{\mu\nu}(\omega')\right]}{\omega'-\omega} \nonumber
    \\
    &= \frac{1}{2} \int_{-\infty}^{+\infty}\frac{d\omega'}{\pi}\left\{\frac{\text{Im}\left[\chi_{\mu\nu}(\omega')\right]}{\omega'-\omega-i\epsilon}+\frac{\text{Im}\left[\chi_{\mu\nu}(\omega')\right]}{\omega'-\omega+i\epsilon}\right\} \label{eq:KrKrRe},
\end{align}
the principal values can be evaluated using contour integration techniques. To do this properly, we first need to specify what branch cuts to use. This will depend on the value of $\omega$. Denote the branch points as $\omega_i$, they form the set $\{\omega_j\}_{j=1}^{2m}$, and are ordered such that $\omega_j<\omega_{j+1}$ for all $j$. Then, we will have the following two different cases:
\begin{enumerate}
    \item If $\omega<\omega_1$, or $\omega_{2m}<\omega$, or $\omega_1<...<\omega_{2a}<\omega<\omega_{2a+1}<...<\omega_{2m}$ for some $1<a<m-1$, the branch cuts are formed between neighboring branch points, i.e., all branch cuts are finite and made between $\omega_{2j-1}$ and $\omega_{2j}$, for $j=1,...,m$.
    \item If $\omega_1<...<\omega_{2a-1}<\omega<\omega_{2a+1}<...<\omega_{2m}$ for some $1<a<m-1$, we have infinite branch cuts on the interval $(-\infty,\omega_1]$ and $[\omega_{2m},+\infty)$, and finite branch cuts between neighboring pairs $\omega_{2j}$ and $\omega_{2j+1}$, for $1<j<m-1$.
\end{enumerate} 
These are schematically depicted with the resulting integration contours in FIG.~\ref{fig:integration contours}.

Knowing where the branch cuts are located, we can now specify what contour to use. 
The first integral in Eqs.~\eqref{eq:KrKrIm} and \eqref{eq:KrKrRe} is evaluated by first surrounding the branch cuts with semi circular arcs of radius $r$ in the upper half plane, the pole at $\omega'=\omega$ with a semi circular arc of radius $r$ in the lower half plane, and then closing the contour with a semi circular arc of radius $R$ in the lower half plane. Deforming this contour, we end up with a series of dogbone or dumbell contours, and potentially a pair of hairpin contours (if the branch cuts to infinity are present) wrapping around the branch cuts, as depicted in FIGs.~\ref{fig:integration contours} (c) and (f). The second integral in Eqs.~\eqref{eq:KrKrIm} and \eqref{eq:KrKrRe} is evaluated by first surrounding the branch points with semi circular arcs of radius $r$ in the upper half plane, the pole at $\omega'=\omega$ with a semi circular arc of radius $r$ in the upper half plane, and then closing the contour with a semi circular arc of radius $R$ in the upper half plane. This contour will vanish completely upon deformation. Using the Residue theorem, and the ML estimation method to show that the contribution from all semi circular arcs around the branch points and from the infinite arc vanish when $r\to 0$, and $R\to \infty$, respectively, we finally arrive at
\begin{widetext}
\begin{align}
    \text{Re}\left[\chi_{xx}(\omega)\right] &= \text{PV}\int_{-\infty}^{\infty} \frac{d\omega'}{\pi}\frac{\text{Im}\left[\chi_{xx}(\omega')\right]}{\omega'-\omega} \nonumber
    \\
    &= \begin{cases} &\sum_{j=1}^{m}\int_{\omega_{2j-1}}^{\omega_{2j}}\frac{d\omega'}{\pi}\frac{\text{Im}\left[\chi_{xx}(\omega')\right]}{\omega'-\omega}, \quad \omega \text{ has even number of poles to the left,}
    \\
    &\int_{-\infty}^{\omega_1}\frac{d\omega'}{\pi}\frac{\text{Im}\left[\chi_{xx}(\omega')\right]}{\omega'-\omega}+\int_{\omega_{2m}}^{+\infty}\frac{d\omega'}{\pi}\frac{\text{Im}\left[\chi_{xx}(\omega')\right]}{\omega'-\omega}+\sum_{j=1}^{m-1}\int_{\omega_{2j}}^{\omega_{2j+1}}\frac{d\omega'}{\pi}\frac{\text{Im}\left[\chi_{xx}(\omega')\right]}{\omega'-\omega}, \text{ otherwise,} \end{cases} \label{eq:PVxx}
    \\
    -\text{Im}\left[\chi_{xy}(\omega)\right] &= \text{PV}\int_{-\infty}^{\infty} \frac{d\omega'}{\pi}\frac{\text{Re}\left[\chi_{xy}(\omega')\right]}{\omega'-\omega} \nonumber
    \\
    &= \begin{cases} &\sum_{j=1}^{m}\int_{\omega_{2j-1}}^{\omega_{2j}}\frac{d\omega'}{\pi}\frac{\text{Re}\left[\chi_{xy}(\omega')\right]}{\omega'-\omega}, \quad \omega \text{ has even number of poles to the left,}
    \\
    &\int_{-\infty}^{\omega_1}\frac{d\omega'}{\pi}\frac{\text{Re}\left[\chi_{xy}(\omega')\right]}{\omega'-\omega}+\int_{\omega_{2m}}^{+\infty}\frac{d\omega'}{\pi}\frac{\text{Re}\left[\chi_{xy}(\omega')\right]}{\omega'-\omega}+\sum_{j=1}^{m-1}\int_{\omega_{2j}}^{\omega_{2j+1}}\frac{d\omega'}{\pi}\frac{\text{Re}\left[\chi_{xy}(\omega')\right]}{\omega'-\omega}, \text{ otherwise.} \label{eq:PVxy} \end{cases}
\end{align}
\end{widetext}
which is the central result of this section.

\subsection{DC-limit} \label{sec:DCexpr}

The expressions derived above are only valid for non-zero frequencies, meaning that the DC-limit ($\omega\to 0$) has to be studied separately. In terms of the response function, this has to be done with caution as the order of various limits and integrations not necessarily commute. Recalling the Kubo formula, the conductivity reads in the DC-limit reads
\begin{equation}
    \sigma_{\mu\nu}(0) = \lim_{\omega\to 0}\left\{\frac{1}{i\omega}\left[\chi_{\mu\nu}(\omega)-\chi_{\mu\nu}(0)\right]\right\} = \frac{1}{i}\frac{d}{d\omega}\chi_{\mu\nu}(\omega)_{\left. \right|_{\omega=0}}.
\end{equation}
Instead of carrying out this derivative, we here choose to employ the Kramers-Kronig relations once again, and illustrate how these can be used to arrive at an analytical closed-form expression for the DC Hall conductivity. This reads
\begin{align}
    \text{Re}\left[\sigma_{xy}(0)\right] &= \lim_{\omega\to 0} \frac{1}{\omega}\left\{\text{Im}\left[\chi_{xy}(\omega)\right]-\text{Im}\left[\chi_{xy}(0)\right]\right\} \nonumber
    \\
    &=\lim_{\omega \to 0}\text{PV}\int_{-\infty}^{+\infty} \frac{d\omega '}{\pi} \frac{\text{Re}\left[\chi_{xy}(\omega')\right]}{\omega'\left(\omega'-\omega\right)}.
\end{align}
Before formally taking the limit, we need to make sure that this can be done in a consistent way with respect to the choice of branch cuts. For this, we assume that $\omega$ is very small, so that there exists no branch points between $\omega'=0$ and $\omega'=\omega$. This corresponds to case 1 in Sec.~\ref{sec:reactcomp}, and thus the DC Hall conductivity reads
\begin{equation} \label{eq:dclimit}
    \text{Re}\left[\sigma_{xy}(0)\right] = -\lim_{\omega \to 0}\sum_{j=1}^m \int_{\omega_{2j-1}}^{\omega_{2j}}\frac{d\omega'}{\pi}\frac{\text{Re}\left[\chi_{xy}(\omega')\right]}{\omega'(\omega'-\omega)}.
\end{equation}
If the limit $\omega\to 0$ is now naively taken before the integration is performed, the pole at $\omega'=0$ will become a pole of order 2 instead of order 1. Consequently, the integration along the small semi-circular contours around $\omega'=0$ will not vanish, but rather result in divergences which not necessarily cancel each other. Therefore, this is as far as the analytical simplifications go, and we will use Eq.~\eqref{eq:dclimit} to evaluate the Hall conductivity in the DC-limit.

\section{Magneto-optical conductivity for inversion symmetric Weyl nodes} \label{sec:results}
We now turn to apply the techniques developed above for concrete setups. We will use a model whose dispersion is given by
\begin{align}
    g(k_z) &:= \gamma \left(k_z^2-\alpha^2\right)\left(k_z^2+\beta^2\right),
    \\
    h(k_z) &:= ak_z^4+bk_z^3+ck_z^2+dk_z+e,
\end{align}
where all the constants are real. Such a system has inversion-symmetric Weyl nodes at $k_z=\pm \alpha$. 
Both $g(k_z)$ and $h(k_z)$ are taken to be fourth order, and the parameters can be tuned such that potential Fermi pockets in connection to the nodes will be kept finite in size.
The coefficients of $h(k_z)$ will be defined through
\begin{equation}
    h(\pm \alpha) = E_{\text{W}_{\pm}}, \quad h'(\pm \alpha) = \eta_{\pm},
\end{equation}
where $E_{\text{W}_{\pm}}$ denote the energy at the Weyl node at $k_z=\pm \alpha$, and $\eta_{\pm}$ controls the tilt at $k_z=\pm \alpha$. Manually setting $a$ to assure finite Fermi pockets, and $E_{\text{W}_{\pm}}=0$, the coefficients of $h(k_z)$ are defined as 
\begin{align}
    a &= 0.01, \label{eq:a}
    \\
    b &= \frac{\eta_++\eta_-}{4\alpha^2},\label{eq:b}
    \\
    c &= \frac{-2b\alpha^2-4\alpha^3+\eta_+}{2\alpha},\label{eq:c}
    \\
    d &= -b\alpha^2,\label{eq:d}
    \\
    e &= -d\alpha-c\alpha^2-b\alpha^3-a\alpha^4.\label{eq:e}
\end{align}
Initially, we will investigate how the conductivity behaves when varying the tilt parameters. For the sake of clarity, we will treat different kinds of systems in separate subsections. We will focus on the real parts of $\sigma_{xx}$ and $\sigma_{xy}$ for the results, i.e., we will display one dissipative and one reactive component of the response function. 

\subsection{Untilted and tilted type I} \label{sec:untilt}

\begin{figure}[hbt!]
\centering

\includegraphics[width=\columnwidth]{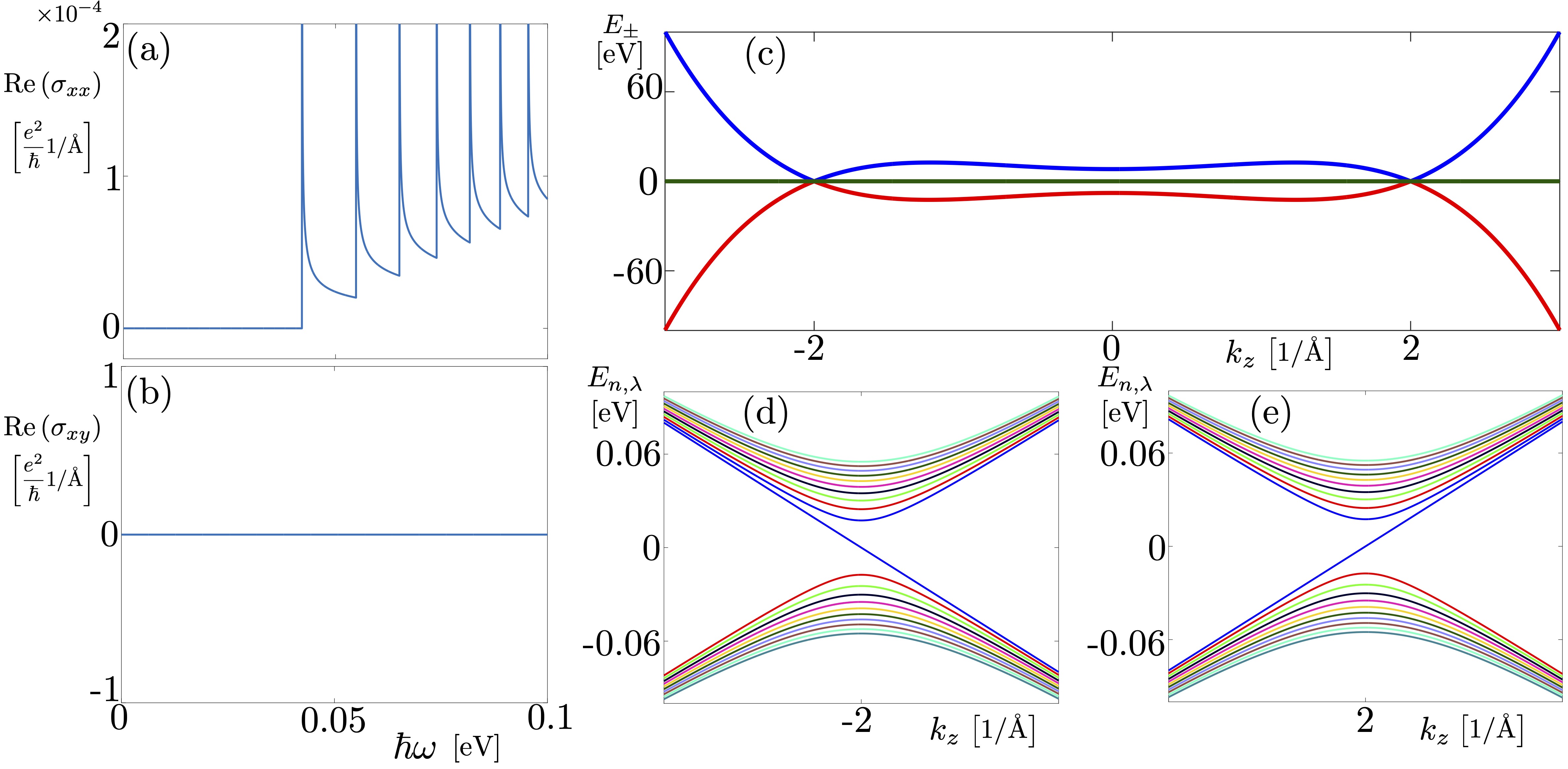}

\caption{Real parts of $\sigma_{xx}(\omega)$ (a) and $\sigma_{xy}(\omega)$ (b) for parameter values $\alpha=2$, $\beta=1$, $\gamma=2$, $\eta_+=\eta_-=0$, $\mu=0$eV and $B=10$T. The band structure in the absence of a magnetic field is shown in (c), while (d) and (e) show the Landau level structure at $k_z=\pm \alpha$. The peak structure in $\sigma_{xx}$ is kept also when two nodes are considered and corresponds to interband transitions between Landau levels, but the contribution from the transition between the chiral level and the first Landau level is cancelled. This is because there will be exactly opposite contributions from the two nodes, one from the chiral level to the first Landau level at one node, and the reversed transition at the second node. Since there are no Fermi pockets, there are no allowed intrarband transitions, causing $\sigma_{xy}$ to be vanishing for all $\omega$. \label{fig:untilteduntilted} }
\end{figure}

Let us start with a system where both nodes are untilted, i.e., where $\eta_+=\eta_-=0$. 
The real parts of $\sigma_{xx}$ and $\sigma_{xy}$ are displayed in FIG.~\ref{fig:untilteduntilted} (a) and (b), respectively, where the $xx$-components displays the characteristic peak-structure, corresponding to optical transitions between neighboring Landau levels of different chirality.
The position of the peaks correspond exactly to the difference in energy between respective Landau level pairs, which agrees with previous studies where typically systems hosting only one Weyl node are studied \cite{AC2013,Tchoumakov2016,Yadav2022,Stalhammar20}. 
An important difference here, however, is that the contribution from the transition from the chiral level is absent. 
This is seen as a small bump located at the energy corresponding to the difference between the chiral level and the first Landau level at the Weyl node, cf. Refs.~\cite{AC2013,Tchoumakov2016,Yadav2022,Stalhammar20}. 
Here, however, we have matching, but opposite, contributions from the two nodes, which makes the net contribution vanish. 
As $\mu=0$ in FIG.~\ref{fig:untilteduntilted}, the Hall conductivity vanish for all $\omega$, and provides nothing interesting at this stage.

\begin{figure}[hbt!]
\centering

\includegraphics[width=\columnwidth]{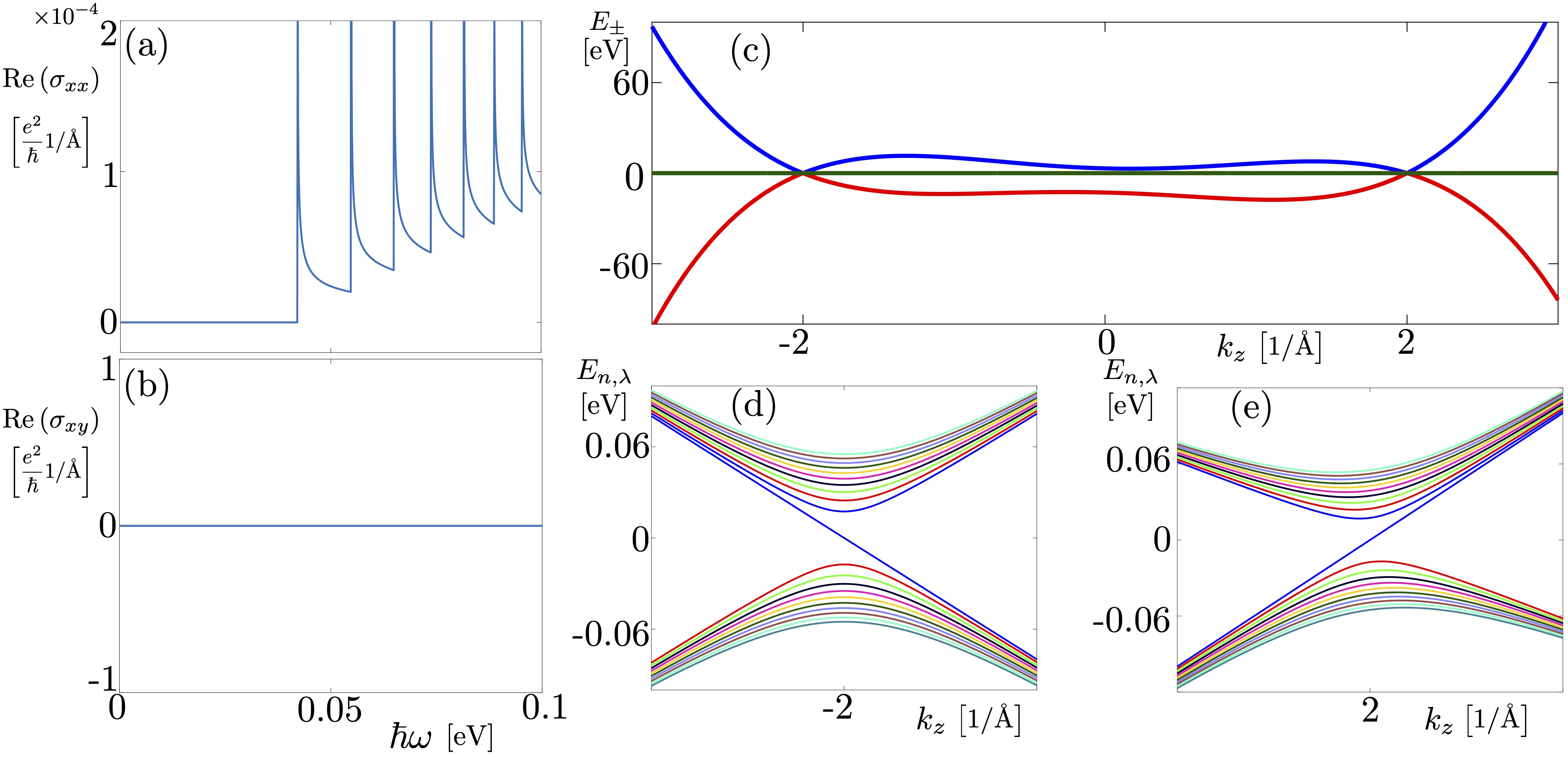}

\caption{Real parts of $\sigma_{xx}(\omega)$ (a) and $\sigma_{xy}(\omega)$ (b) for parameter values $\alpha=2$, $\beta=1$, $\gamma=2$, $\eta_+=10$, $\eta_-=0$, $\mu=0$eV and $B=10$T. The band structure in the absence of a magnetic field is shown in (c), while (d) and (e) show the Landau level structure at $k_z=\pm \alpha$. The weak tilting of the node at $k_z=2$ leaves the behaviour of both $\sigma_{xx}$ and $\sigma_{xy}$ unchanged, which is a consequence from there still being no allowed intraband transitions. \label{fig:typeIuntilted} }
\end{figure}

In FIG.~\ref{fig:typeIuntilted} the dispersion around on of the nodes have been tilted slightly, while the other is kept untilted. 
As the tilt is weak enough to ensure that neither the valnce band nor the conduction band cross, but merely touch, the Fermi level at the Weyl nodes, there are no additional transitions that can occur. Therefore, the result in this case very much resembles that of two untilted nodes; the $xx$-component still has its characteristic peaks, while the $xy$–component remains vanishing.

\subsection{Tilted type II} \label{sec:typeII}
\begin{figure}[hbt!]
\centering

\includegraphics[width=\columnwidth]{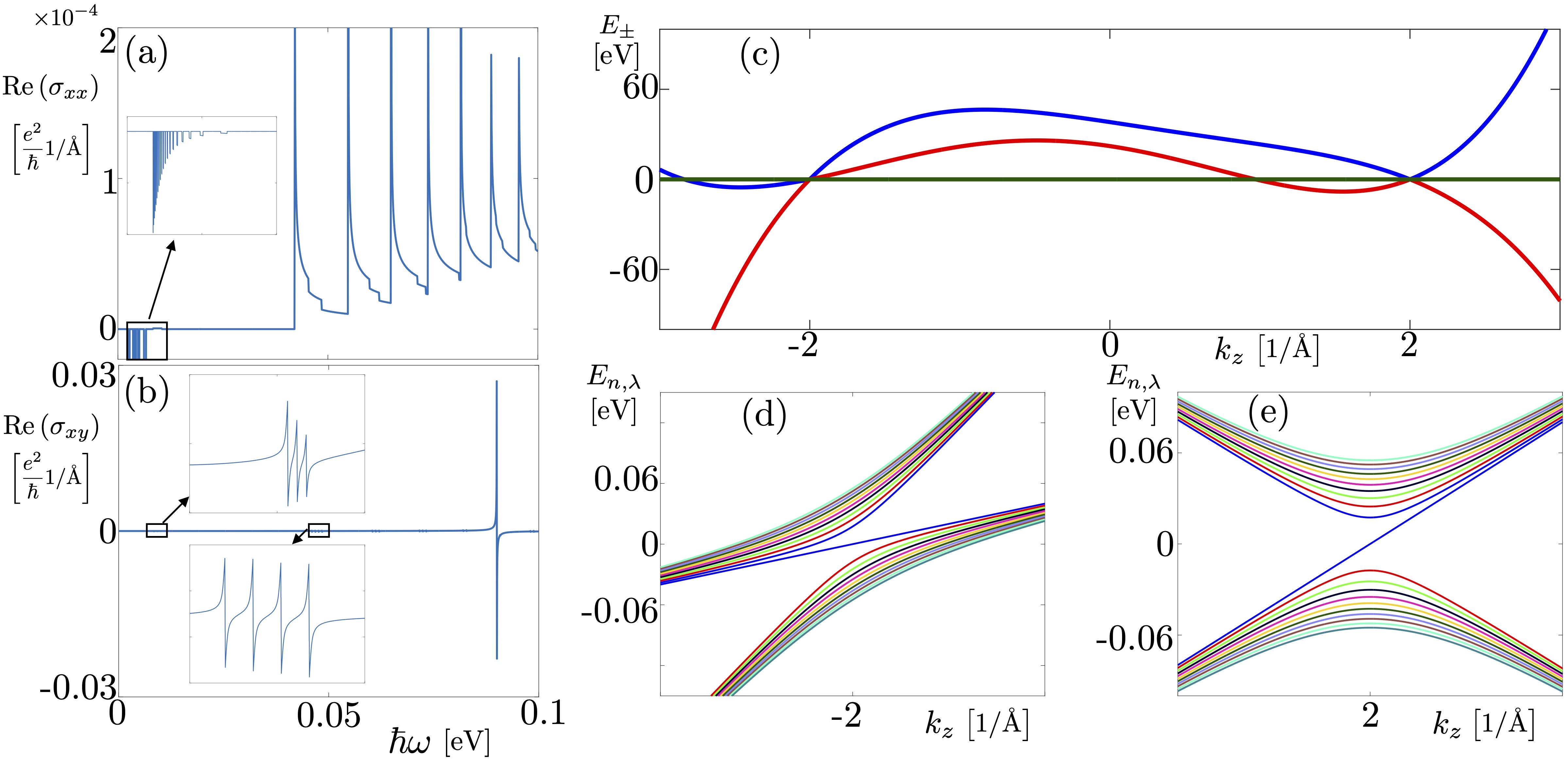}

\caption{Real parts of $\sigma_{xx}(\omega)$ (a) and $\sigma_{xy}(\omega)$ (b) for parameter values $\alpha=2$, $\beta=1$, $\gamma=2$, $\eta_+=0$, $\eta_-=60$, $\mu=0$eV and $B=10$T. The band structure in the absence of a magnetic field is shown in (c), while (d) and (e) show the Landau level structure at $k_z=\pm \alpha$. As a consequence of the existence of Fermi pockets, intraband transitions are allowed, which gives a response at smaller frequencies. Additionally, there are interband transitions happening at non-zero $k_z$, which can be seen as the small steps between the peaks in panel (a). These transitions do not happen in a symmetric fashion, since the Fermi pockets are only present in direct connection to the tilted node. This gives a response also in the Hall conductivity, which is seen as sharp transitions in panel (b). Mathematically, these correspond exactly to at what frequencies the choices of branch cuts, and hence integration contours, are changed, making it possible to interpret the Hall response as a branch point transition.
 \label{fig:untiltedtypeII} }
\end{figure}

When the tilt is increased further, the valence and/or conduction band may fully cross the Fermi level, and the Weyl semimetal enter the type-II-phase. 
This phase is distinguished from the type-I phase by that the Fermi surface becomes finite and thus is no longer point-like. 
In the context of conductivity, this allows for intraband transitions, i.e., transitions between neighboring Landau levels of the same chirality. 
Just as for the interband transitions, they appear at energies corresponding exactly to the energy difference the respective Landau level pairs, which in this case are spaced much closer together. 
This can be seen in FIG.~\ref{fig:untiltedtypeII}, where one of the nodes are kept untilted and the other is tilted such that the Fermi pockets are formed, cf. FIG.~\ref{fig:untiltedtypeII}(c)-(e). 
In addition to the interband peaks, the intraband peaks are located close to $\hbar\omega=0$. 
Notably, the type-II phase also host a non-trivial Hall conductivity $\sigma_{xy}$, which can be seen in FIG.~\ref{fig:untiltedtypeII}(b).

\begin{figure}[hbt!]
\centering

\includegraphics[width=\columnwidth]{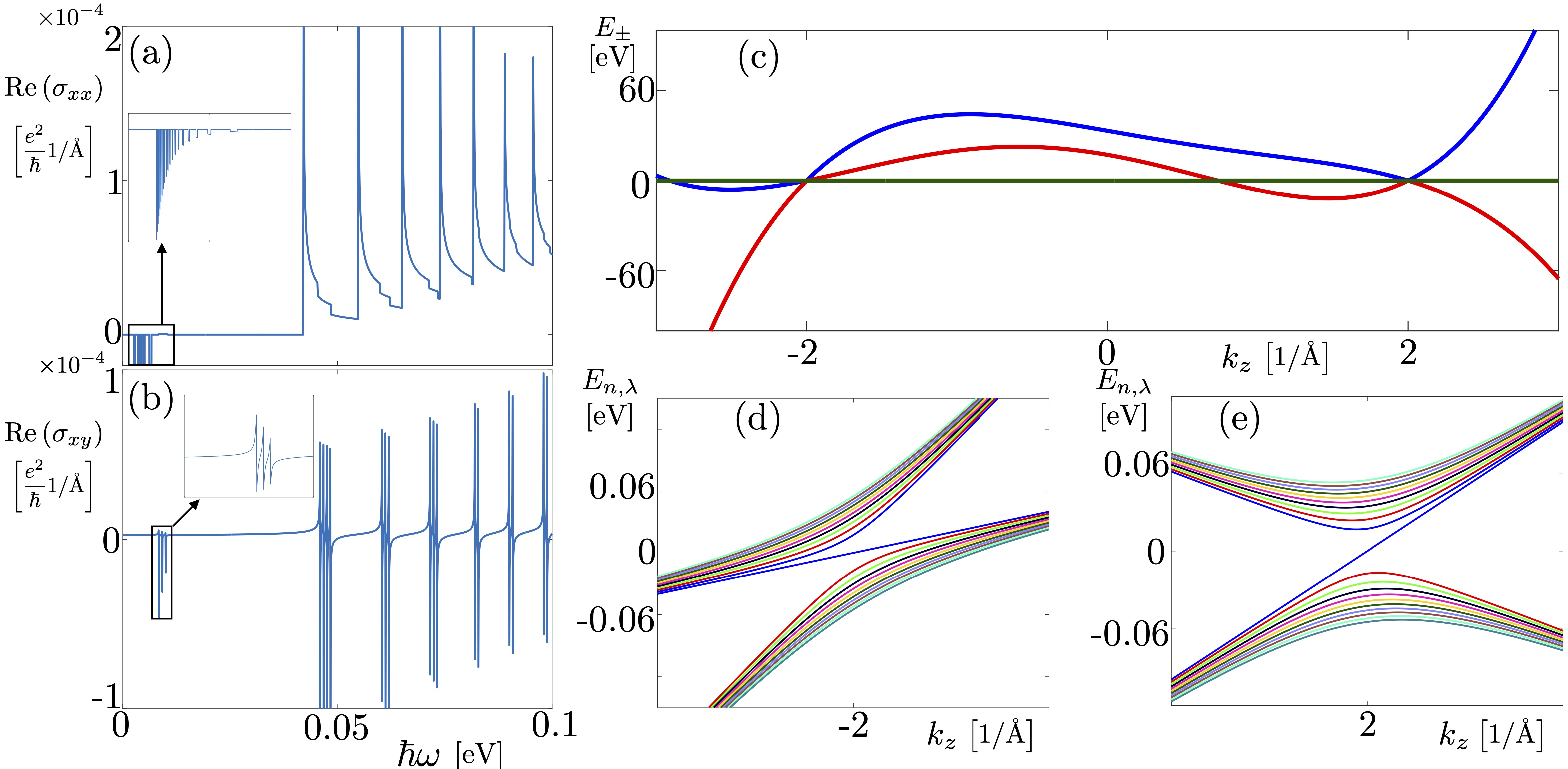}

\caption{Real parts of $\sigma_{xx}(\omega)$ (a) and $\sigma_{xy}(\omega)$ (b) for parameter values $\alpha=2$, $\beta=1$, $\gamma=2$, $\eta_+=10$, $\eta_-=60$, $\mu=0$eV and $B=10$T. The band structure in the absence of a magnetic field is shown in (c), while (d) and (e) show the Landau level structure at $k_z=\pm \alpha$. The weak tilting of the node located at $k_z=2$ does not affect the result significantly from the case where this node is untilted (cf. FIG.~\ref{fig:untiltedtypeII}).The peak structure in (a) behaves similarly, and we see essentially equivalent behavior of the Hall conductivity in (b), even though it seems as if the strength of the response is more even throughout the considered energy spectrum. 
This could, however, just be a consequence of the evaluation of the principal value---the size of the peaks are not to be taken literal. 
  \label{fig:typeItypeII} }
\end{figure}

When just slightly tilting the previously untilted node, noting interesting initially happens (cf. FIG.~\ref{fig:typeItypeII}), but when the second node eventually is overtilted as well, the behaviour of the conductivity changes significantly. 
Notably, the Hall conductivity becomes highly oscillating along all energies shown in FIG.~\ref{fig:typeItypeII}, while the $xx$-component is left somewhat unchanged, with the importance exception being a slight deviation from 0 at the same frequencies as the oscillatory behavior of $\sigma_{xy}$ begins. 
This behavior is caused by the existence of two particle-hole pairs of Fermi pockets, vastly increasing the number of allowed Landau level transitions.

\begin{figure}[hbt!]
\centering

\includegraphics[width=\columnwidth]{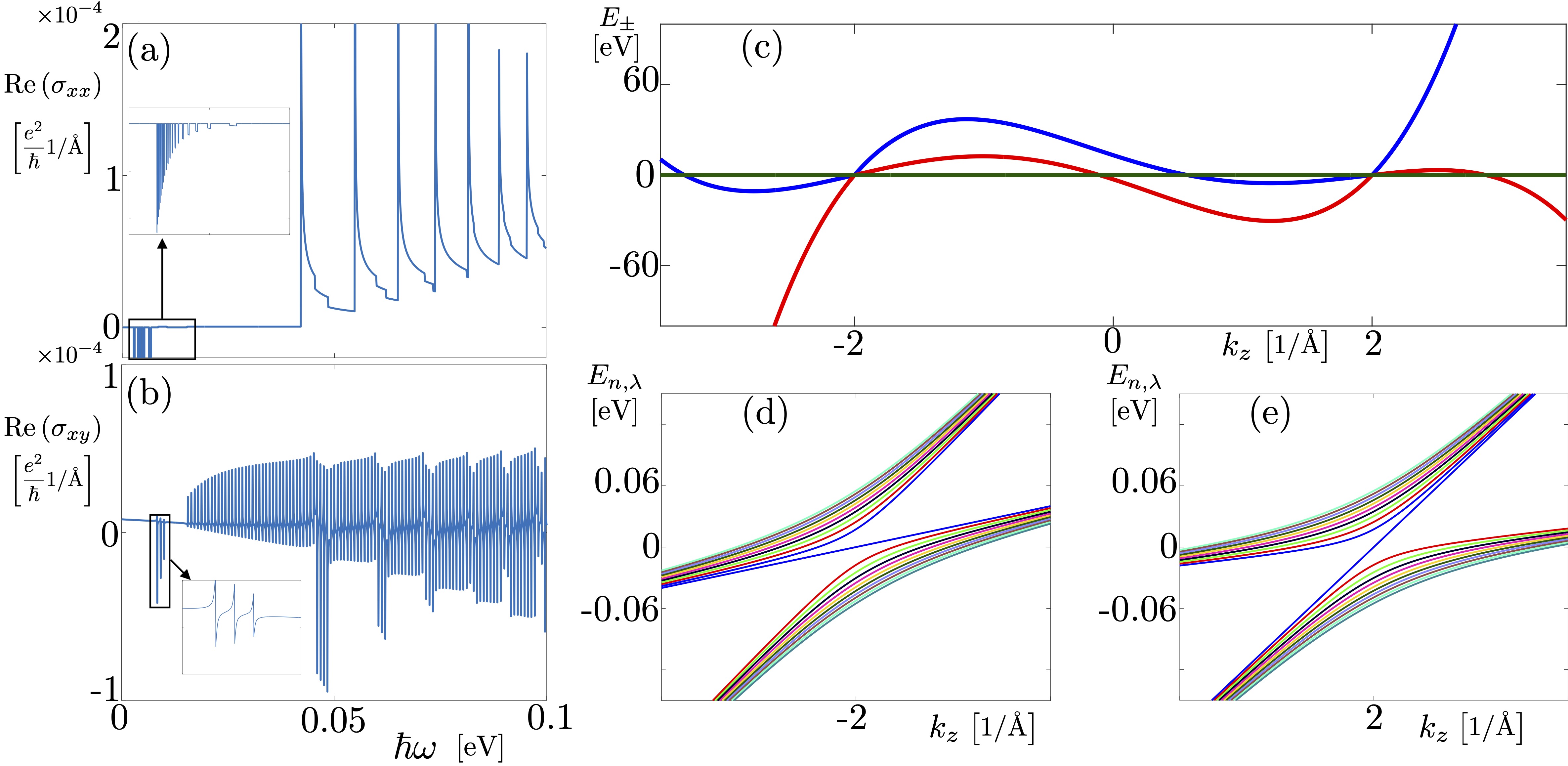}

\caption{Real parts of $\sigma_{xx}(\omega)$ (a) and $\sigma_{xy}(\omega)$ (b) for parameter values $\alpha=2$, $\beta=1$, $\gamma=2$, $\eta_+=50$, $\eta_-=60$, $\mu=0$eV and $B=10$T. The band structure in the absence of a magnetic field is shown in (c), while (d) and (e) show the Landau level structure at $k_z=\pm \alpha$. The behaviour of $\sigma_{xx}$ seems to remain the same as in FIG.~\ref{fig:typeItypeII}, but the response of the Hall conductivity has become highly oscillatory. Taking a closer look, $\sigma_{xx}$ actually deviates slightly from 0 starting at the same frequency as the oscillations of $\sigma_{xy}$. These two features indicate that the number of allowed transitions vastly increases in this regime, which can be explained by the existence of two particle-hole pairs of Fermi pockets in the band structure (a).
   \label{fig:typeIItypeII} }
\end{figure}
\subsection{Trivial pockets} \label{sec:trivialpockets}

\begin{figure}[hbt!]
\centering

\includegraphics[width=\columnwidth]{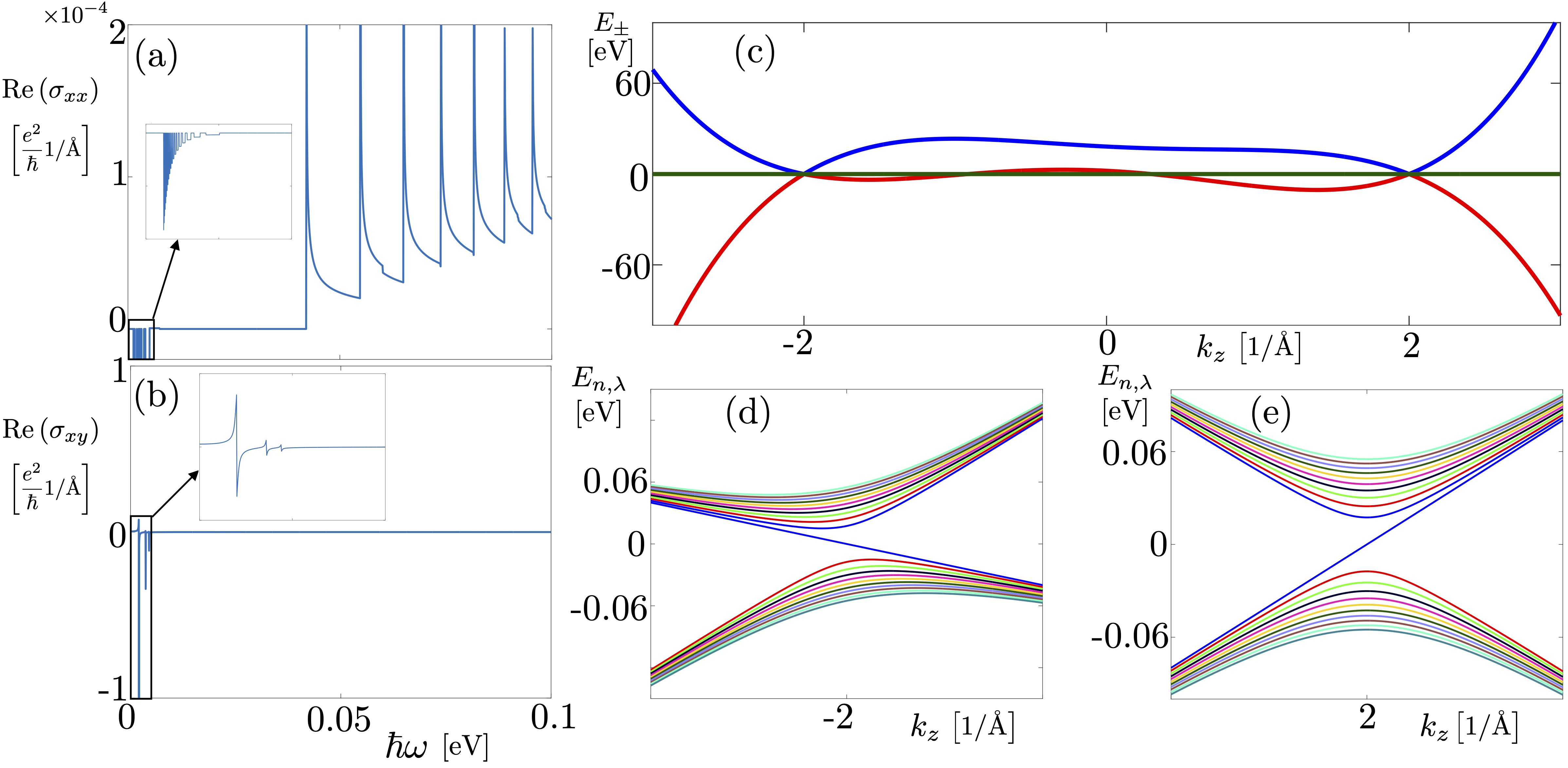}

\caption{Real parts of $\sigma_{xx}(\omega)$ (a) and $\sigma_{xy}(\omega)$ (b) for parameter values $\alpha=2$, $\beta=1$, $\gamma=2$, $\eta_+=0$, $\eta_-=20$, $\mu=0$eV and $B=10$T. The band structure in the absence of a magnetic field is shown in (c), while (d) and (e) show the Landau level structure at $k_z=\pm \alpha$. The presence of trivial Fermi pockets result in a response similar to those features observed in type II systems, which is a consequence of intraband transitions being allowed. This results in a Hall response (b), and a sharp contribution for very low frequencies for the $xx$-component (a). \label{fig:trivialpockets} }
\end{figure}
Another interesting phenomena that might be present in Weyl semimetals, is trivial Fermi pockets. 
These are formed when any of the bands cross the Fermi energy away from a Weyl node. 
This also allows for intraband transitions, and these systems are therefore very much reminiscent of a type-II-system. 
In FIG.~\ref{fig:trivialpockets} we display the conductivity for such a system, which is created by tilting one of the nodes slightly, but not into the type-II regime. 
As a consequence of the existence of trivial Fermi pockets and the allowed intraband transitions, the conductivity very much behaves like that in a type-II system.

\section{Discussion and Experimental Relevance of Results} \label{sec:discussion}
The results in Sec.~\ref{sec:results} show that studying systems with a chiral pair of Weyl nodes gives two important differences compared to those including only one node; the contribution from the transition to/from the chiral level at the Weyl nodes exactly cancel each other, and the frequencies at which intraband transitions occur are smeared out. This claim is supported by comparing the systems studied in Sec.~\ref{sec:results} to those studied in, e.g., Ref.~\cite{Stalhammar20}, which display intraband contributions at significantly higher frequencies at similar field strengths. 
This serve as an indicator that additional nodes smears out the energy scale at which the intraband transitions are allowed, and that considering even more nodes could be a way to understand why the distinct peak structure sourced by interband transitions is not visible in experiments; if intraband and interband transitions occur at similar energy scales, the response from the intraband transitions will, due to their significantly larger response signal, mask that of the interband transitions and provide a possible explanation for the mismatch between theory and experiments. 
The absent of a clear chiral contribution is due to the fact that the energy splittings between the chiral level and the first Landau level are equal exactly at the Weyl nodes. It is important to note that this does not completely eliminate the contribution from the chiral level, since transitions away from the Weyl nodes may very well occur, but the strength of these are highly suppressed by competing transitions (cf. FIG.~\ref{fig:typeIItypeII}).
The individual signatures from both intraband and interband transitions are, however, remained intact when higher-order terms of momentum and additional Weyl nodes are considered, which indicate that to study these phenomena separately, systems hosting only one Weyl node will suffice. What studies beyond single node systems contribute with is the important interplay between these signature, that could serve as an explanation for the mismatch between theory and experiment.
This is of course not the only such possible source, and we now turn to discuss some of them separately.

\subsection{Beyond the Clean Limit} \label{sec:cleanlimit}
All the results presented in Sec.~\ref{sec:results} display a very sharp peak structure, something that is not seen in experiments \cite{PD2020,XZC2018,PGW2020}. 
This structure is (theoretically) present regardless of if the approximation of the studied Weyl systems is linear \cite{AC2013,Tchoumakov2016}, considers only one node \cite{Stalhammar20,Yadav2022}, or, as is shown in this work, if a pair of nodes is considered. 
This particular mismatch is most likely a consequence of the theoretical studies being carried out in the clean limit, i.e., the regulator $\epsilon$ in Eq.~\eqref{eq:linres} is taken to zero. 
A finite regulator can be thought of as assigning impurities, or, more concretely, a finite scattering life time in the system \cite{AC2013}. 
This will slightly smear out the peaks, and make the overall structure of the conductivity smoother. 
Such numerical simulations however are beyond the current scope and the exact impact of impurity scattering is left for future studies. 

\subsection{Symmetry-Breaking and Nodes at Different Energies} \label{sec:symmbreak}

\begin{figure}[hbt!]
\centering

\includegraphics[width=\columnwidth]{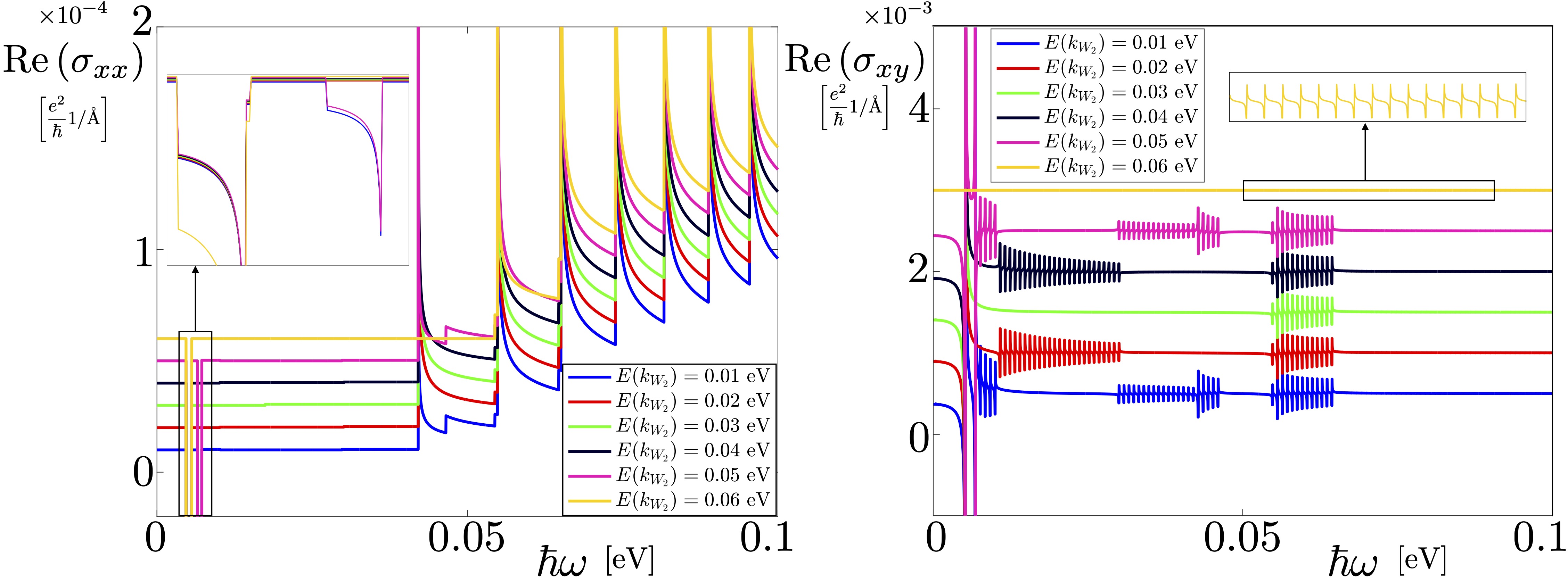}

\caption{Real parts of $\sigma_{xx}(\omega)$ (a) and $\sigma_{xy}(\omega)$ (b) for parameter values $\alpha=2$, $\beta=1$, $\gamma=2$, $\eta_+=0$, $\eta_-=20$, and $B=10$T, for different location in energy of the Weyl node at $k_z=-\alpha$. The results for different such energies are shifted in steps of $10^{-5}$ in (a) and $2*10^{-4}$ in (b) to allow reasonable visibility. The conductivity properties changes qualitatively with the energy difference between the nodes, as it allows for some additional transitions, and forbids some that was previously allowed. For instance, the first interband transition for $\mu=0.06$eV is not the same as for the other values of $\mu$, which can be seen in (a) as the first peak of the yellow curve coincides with the second peak of the other curves. Having a non-zero energy difference between the nodes also gives a Hall response, seen in (b), but this time as a consequence of allowed intraband and interband transitions. The former further gives a sharp response at lower frequencies in (a).    
 \label{fig:diffenergies} }
\end{figure}

In this work, we have restricted ourselves to present plots of systems where the Weyl nodes are spaced symmetrically in momentum space, and where they both appear at zero energy. It should be noted that the general reasoning, and hence the expressions for the linear response function is applicable to any general polynomials $g(k_z)$ and $h(k_z)$, and that the restriction is only in terms of the displayed plots. However, as the behaviour of both $\sigma_{xx}$ and $\sigma_{xy}$ can be thoroughly understood in terms of Landau level transitions, this restriction can be made almost without loss of any generality. This can be understood as follows.

Consider first the case where the nodes are not inversion symmetric. Regardless of at what momenta they are located, the Landau levels at the Weyl nodes will be equally split in energy---this splitting is not depending on the position of the nodes in momentum space, but rather by the magnetic field [recall that they appear exactly when $g(k_z)=0$]. The position of the nodes might however affect the Landau level splitting at the points where the intraband transition occurs, as these generically happen at momenta where $g(k_z)$ is non-zero. But this is also the case for inversion symmetric nodes, as the tilt at the respective nodes will source similar features in the Landau level splitting. Hence, breaking inversion symmetry is not expected to give rise to any new physics, but will rather change the location of the intraband transitions in a way similar to what the tilting at the respective nodes does.

Now, what would happen if the nodes appear at different energies? Again, this will not change the Landau level splitting at the nodes, but it will change the allowed transitions and break the symmetry between the transitions among the nodes. This would allow for contributions from interband transitions to the Hall conductivity, illustrated in FIG.~\ref{fig:diffenergies}, along with the more conventional contributions from intraband transitions. Thus, it can be concluded that the location of the nodes in energy space is reflected in the optical response.

\subsection{DC-limit}\label{sec:dcres}

\begin{figure}[hbt!]
\centering

\includegraphics[width=\columnwidth]{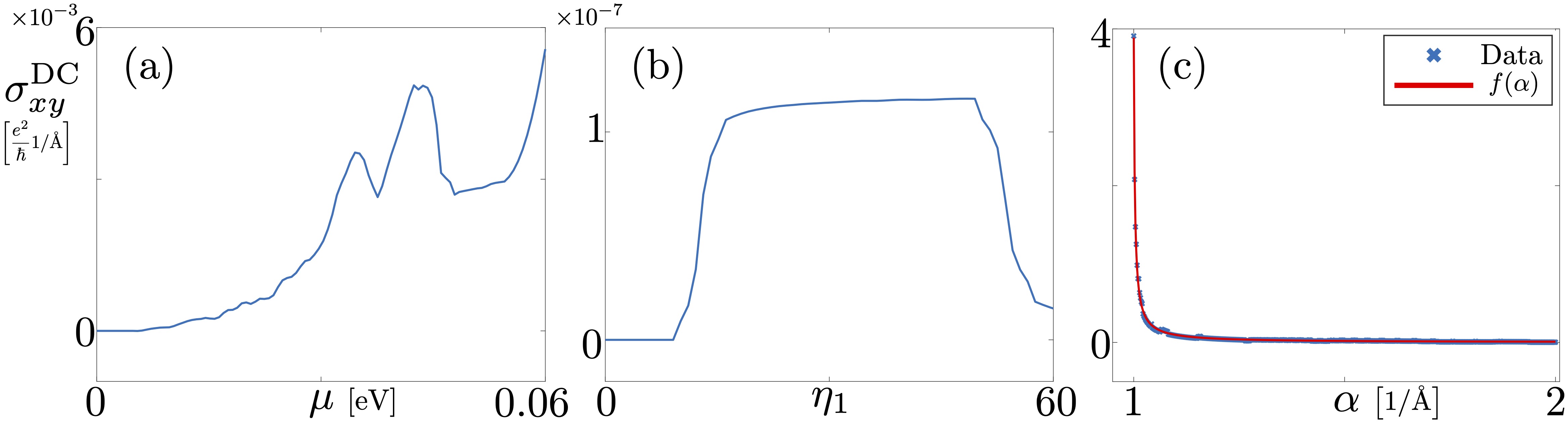}

\caption{DC Hall conductivity as a function of chemical potential $\mu$ (a), tilting at the node at $k_z=\alpha$ (b), and the splitting between the nodes (c) for $\alpha=2$ (a), (b), $\beta=1$, $\gamma=2$, $\eta_+=0$, $\eta_-=20$ (a), (c), $\mu=0.03$eV (b), (c) and $B=10$T. Albeit being an unstable expression, Eq.~\eqref{eq:dclimit} captures the gluing features predicted for (a) and (b), i.e., that the Hall conductivity should consist of piecewise smooth functions of chemical momentum or tilt. In (c), it becomes apparent that the distance between the Weyl nodes is highly relevant for the strength of the response signal. As a function of the location of the respective Weyl nodes, the conductivity can be approximated as $\sigma_{xy}^{\text{DC}}(\alpha) = \frac{c_1}{\alpha^3+c_2\alpha^2+c_3\alpha+c_4}$, with $(c_1,c_2,c_3,c_4) = 10^3\cdot(0.0224,0.2516,1.7408,-1.9876)$. \label{fig:DCHall} }
\end{figure}

Albeit the analytical tools used in this work are not ideal in terms of stability for investigating the DC-limit, we will devote this subsection to understanding the DC Hall conductivity as a function of various system parameters. 
But before evaluating Eq.~\eqref{eq:dclimit} explicitly, let us think about what conclusions can be made from the expression itself, highlighting the strength of analytical frameworks when it comes to phenomenological understanding. 
We know that what contributes to the optical Hall conductivity are optical Landau level transitions, which due to the Kramers-Kronig relations, Eqs.~\eqref{eq:KrKrIm} and \eqref{eq:KrKrRe}, are also what will contribute in the DC-limit. 
This means that we should observe a step-wise change in the DC-conductivity when changing the chemical potential, since this will control the number of allowed transitions. 
When $\mu$ sweeps through an additional Landau level, the curve will change shape. 
However, it does not necessarily mean that we expect a clean plateau structure, not even in the $T\to 0$-limit, even though the Fermi functions become step-functions. 
Instead, the effect is that the integration domain is extended, and therefore the DC Hall conductivity should consist of a series of glued, piecewise smooth functions. 
This is exactly what is observed in Ref.~\cite{Xiong2022}. 
Interestingly enough, the authors of Ref.~\cite{Xiong2022} are understanding this behaviour in terms of the density of states, while our reasoning connects the same phenomena to Landau level transitions and choices of branch cuts. 
We include a plot of the DC Hall conductivity as a function of $\mu$ in FIG.~\ref{fig:DCHall} (a), obtained from evaluating Eq.~\eqref{eq:dclimit}. 

The chemical potential is not alone responsible for controlling which intraband transitions that are allowed. The tilts at the different nodes has a similar effect, meaning that the piecewise structure should be observed also as a function of $\eta_+$ and $\eta_-$ in Eqs.~\eqref{eq:a}-\eqref{eq:b}. This is indeed what we see in FIG.~\ref{fig:DCHall} (b). That the tilt has significant impact on the Hall conductivity is something that is also mentioned in Ref.~\cite{Xiong2022}, even though they are merely considering a couple of different tilting values. 

Another feature of the Hall conductivity is its relation to the spacing between the Weyl nodes. 
From field theoretical techniques, it is predicted that the current related to the Hall conductivity is proportional to this distance (in momentum space) in the presence of an electric field \cite{HQ2013}. 
Since the setup we are considering in this work explicitly excludes an electric field, testing this feature is beyond the current scope, but we can instead investigate how the distance between the nodes affect the Hall conductivity in the presence of a constant magnetic field. 
This is shown in FIG.~\ref{fig:DCHall} (c), from which we can conclude that the distance plays a significant role to the Hall response, but the scaling between the two quantities are not very direct. 
In fact it takes a rational polynomial  on the form $\sigma_{xy}^{\text{DC}}(\alpha) = \frac{c_1}{\alpha^3+c_2\alpha^2+c_3\alpha+c_4}$, with $(c_1,c_2,c_3,c_4) = 10^3\cdot(0.0224,0.2516,1.7408,-1.9876)$ to fit the data reasonably, cf. FIG.~\ref{fig:DCHall} (c).

\subsection{Finite Temperature}\label{sec:fintemp}
As all the general and illustrated results are carried out in the limit of zero temperature, the effect of going to finite temperature must be mentioned. 
We will exclusively focus on the temperature dependence of the Hall conductivity, as it has previously been noted that the peak structure in the $\sigma_{xx}$-component is not notably affected at finite temperatures \cite{Stalhammar20}. 
For the Hall conductivity, going to finite temperature will result in additional Matsubara poles in the lower half complex $\omega$-plane. These will be at values of $\omega$ satisfying,
\begin{equation} \label{eq:Matsubara}
    \frac{E_{n,\lambda}(k_i)-\mu}{k_BT}=i\pi(2m-1),
\end{equation}
where $m\leq 0$ to maintain the analytical properties of the response function in the upper half complex $\omega$ plane. Thus, in addition to retaining the original pre-factors $A_n^{\lambda\lambda'}(k_i)$ in terms of the Fermi distribution functions, there will be contributions on the form 
\begin{equation}
   2\pi i \sum_j\text{Res}\left\{\text{Re}\left[\chi_{xy}(\omega)\right],\omega_j\right\},
\end{equation}
where $\omega_j$ solves Eq.~\eqref{eq:Matsubara}. These corrections are not treatable using the analytical tools developed in this work, and hence lie beyond the current scope. 
Their complete contribution amounts to an infinite sum, the convergence and decay of which has to be properly investigated in order to assert a termination of the sum to a finite one to be motivated.
In the DC-limit, and assuming a convergent thermal contribution, one could guess the impact of these corrections. 
As a function of chemical potential, or tilting parameters, the piecewise gluing of curves is expected to be smoothed out when the Fermi functions are re-introduced. 
For low temperatures, the impact of the Matsubara poles is further expected to be small, but for increasing temperatures, these might have a more profound effect that should be investigated in future works.

It should lastly be emphasized that the Kramers-Kronig relations rely on causality, which is what causes the response function to be analytical in the upper half complex $\omega$-plane. 
When considering systems at finite temperature, thus going beyond the zero-temperature limit, the very notion of causality becomes affected, which in turn alters the Kramers-Kronig relations. 
Since the general method calculating the Hall conductivity developed in this work is based exactly on the Kramers-Kronig relations, it is not unexpected that things become complicated when thermal effects are taken into account. 
Exactly how complicated this becomes, and how profound the effects are, is left as an open question to be answered in subsequent studies.

\section{Summary of results and outlook} \label{sec:summary}
In this work, we have expanded on the analytical and general theory calculating magneto-optical conductivity in Weyl semimetals. 
To complement the previous works where the dissipative components of the linear response function have been computed, we here also provide analytical closed-form expressions for the reactive components in terms of a sum of finite integrals (an potentially semi-infinite integrals). 
This allows for the analytical calculation of the optical Hall conductivity beyond the high-field quantum- or DC-limit for generic Weyl semimetals, and for understanding of physical features directly by studying the corresponding band structure in combination with the analytical expressions. 
The theoretical and abstract calculations are then applied to a Weyl system hosting two nodes with potential tilts, which further extends previous works where systems hosting only one node are typically considered. 
Importantly, the simulation includes systems where fourth-order momentum terms are taken into account to make sure that potential Fermi pockets are finite in size. 
This allows us to study magneto-optical transport in principle including contributions from all Landau levels as a function of tilting at both nodes, Weyl node splitting in momentum space, and chemical potential, without the needs of unphysical cut-offs in momentum or energy. 
Apart from the AC optical conductivity, a closed-form expression also for the DC-limit of the Hall conductivity is derived, from which phenomenological conclusions matching with existing literature are made. 

Our study has further verified that neither higher-order corrections in momentum, nor the existence of additional Weyl nodes, changes individual signatures of the magneto-optical spectrum significantly---the individual features linked to interband and intraband transitions seen in the simulations in this work can also be studied in simpler systems hosting only one node. 
There are, however, two important deviations, the first being the absence of a contribution from a transition involving the chiral Landau level, as opposite transitions occur at charge-conjugate Weyl nodes. 
The second, and more important, deviation, is that the relative energy range at which intraband transition occurs seems to be significantly larger in systems with two nodes, than in those with only one node. 
This further strengthens the hypothesis that these transitions, that comes with a significantly larger response signal, are the main reason for the predicted sharp peak structure sourced by interband transitions to be masked in experiments; if intraband transitions are allowed at the same energy scales as interband transitions, the signal from the intraband transition will dominate. 
The general tools developed in this and previous works \cite{Stalhammar20,Yadav2022} can in principle be used to scan through various models and continue the search for the source of this mismatch.

A natural extension of this work would be to generalize the calculational tools to include higher-order Weyl nodes, which in recent works have shown to impact the tilt structure in the optical response \cite{Yadav2022}. 
Additional straight-forward, yet involved, continuations include the finite temperature picture, and the impact of the Matsubara poles in Weyl systems and the corresponding Hall conductivity. 
Lastly, a concrete direction towards finding the source of the mismatch between theory and experiment, would be to use the developed analytical techniques for more realistic band structures of actual materials, and to include interaction effects beyond linear response. 
This would require a combination of experimental measurement techniques, such as ARPES, and first-principle calculations, such as DFT, and comprise a promising path towards unifying experimental observations and theoretical predictions.

\section*{Acknowledgements}
I would like to the group of Prof.~Annica Black-Schaffer, especially Patric Holmvall and Rodrigo Arouca, for discussions and feedback on the results and on the project in general. I would also like to thank Prof.~Thors Hans Hansson, Prof.~Emil J. Bergholtz, and Prof.~Johannes Knolle for discussions and useful comments on the manuscript.

\appendix
\begin{widetext}

\section{Response Function for Weyl semimetals} \label{appendix}
\subsection{Simplifications in the $T\to 0$ limit}
Take as starting point Eqs.~(A26) and (A27) in Ref.~\cite{Stalhammar20}, which state that 
\begin{align}
    &\text{Im}\left[\chi_{xx}(\omega)\right]=-\frac{e^2v_F^2}{16\pi l^2_B}\sum_{n=0}^{n_{\text{max}}}\sum_{i=1}^{2m} \nonumber
    \\
    &\left\{ A_n^{++}(k_i)\left[B^+_n(k_i)B^-_{n+1}(k_i)C_{n,n+1}^{++}(k_i)\theta(-\omega)\theta(\frac{2v_F^2}{l^2_B}-\omega^2)-B_n^+(k_i)B_{n+1}^-(k_i)C^{++}_{n+1,n}(k_i)\theta(\omega)\theta(\frac{2v_F^2}{l^2_B}-\omega^2)\right]\right.\nonumber
    \\
    &+A^{--}_n(k_i)\left[B_n^-(k_i)B^+_{n+1}(k_i)C^{--}_{n,n+1}(k_i)\theta(\omega)\theta(\frac{2v_F^2}{l^2_B}-\omega^2)-B_n^-(k_i)B^+_{n+1}(k_i)C^{--}_{n+1,n}(k_i)\theta(-\omega)\theta(\frac{2v_F^2}{l^2_B}-\omega^2)\right] \nonumber
    \\
    &+A^{+-}_n(k_i)\left[B_n^+(k_i)B^+_{n+1}(k_i)C^{+-}_{n,n+1}(k_i)\theta(\omega)\theta(\omega^2-\frac{2v_F^2}{l^2_B})-B_n^+(k_i)B^+_{n+1}(k_i)C^{-+}_{n+1,n}(k_i)\theta(-\omega)\theta(\omega^2-\frac{2v_F^2}{l^2_B})\right]\nonumber
    \\
    &\left.+A^{-+}_n(k_i)\left[B_n^-(k_i)B^-_{n+1}(k_i)C^{-+}_{n,n+1}(k_i)\theta(-\omega)\theta(\omega^2-\frac{2v_F^2}{l^2_B})-B_n^-(k_i)B^-_{n+1}(k_i)C^{+-}_{n+1,n}(k_i)\theta(\omega)\theta(\omega^2-\frac{2v_F^2}{l^2_B})\right] \right\},
    \end{align}
    \begin{align}
    &\text{Re}\left[\chi_{xy}(\omega)\right]=-\frac{e^2v_F^2}{16\pi l^2_B}\sum_{n=0}^{n_{\text{max}}}\sum_{i=1}^{2m} \nonumber
    \\
    &\left\{ A_n^{++}(k_i)\left[B^+_n(k_i)B^-_{n+1}(k_i)C_{n,n+1}^{++}(k_i)\theta(-\omega)\theta(\frac{2v_F^2}{l^2_B}-\omega^2)+B_n^+(k_i)B_{n+1}^-(k_i)C^{++}_{n+1,n}(k_i)\theta(\omega)\theta(\frac{2v_F^2}{l^2_B}-\omega^2)\right]\right.\nonumber
    \\
    &+A^{--}_n(k_i)\left[B_n^-(k_i)B^+_{n+1}(k_i)C^{--}_{n,n+1}(k_i)\theta(\omega)\theta(\frac{2v_F^2}{l^2_B}-\omega^2)+B_n^-(k_i)B^+_{n+1}(k_i)C^{--}_{n+1,n}(k_i)\theta(-\omega)\theta(\frac{2v_F^2}{l^2_B}-\omega^2)\right] \nonumber
    \\
    &+A^{+-}_n(k_i)\left[B_n^+(k_i)B^+_{n+1}(k_i)C^{+-}_{n,n+1}(k_i)\theta(\omega)\theta(\omega^2-\frac{2v_F^2}{l^2_B})+B_n^+(k_i)B^+_{n+1}(k_i)C^{-+}_{n+1,n}(k_i)\theta(-\omega)\theta(\omega^2-\frac{2v_F^2}{l^2_B})\right]\nonumber
    \\
    &\left.+A^{-+}_n(k_i)\left[B_n^-(k_i)B^-_{n+1}(k_i)C^{-+}_{n,n+1}(k_i)\theta(-\omega)\theta(\omega^2-\frac{2v_F^2}{l^2_B})+B_n^-(k_i)B^-_{n+1}(k_i)C^{+-}_{n+1,n}(k_i)\theta(\omega)\theta(\omega^2-\frac{2v_F^2}{l^2_B})\right] \right\},
\end{align}
with
\begin{align}
	A^{\pm\pm}_n(k_i) &:= f\left[E_{n,\pm}(k_i)\right]-f\left[E_{n+1,\pm}(k_i)\right]=\frac{\sinh{\frac{E_{n+1,\pm}(k_i)-E_{n,\pm}(k_i)}{2k_B T}}}{\cosh{\frac{E_{n+1,\pm}(k_i)+E_{n,\pm}(k_i)-2\mu}{2k_B T}}+\cosh{\frac{E_{n+1,\pm}(k_i)-E_{n,\pm}(k_i)}{2k_B T}}},
	\\
	B^{\pm}_n(k_i) &:= 1\pm\frac{g(k_i)}{\sqrt{g^2(k_i)+\frac{2n}{l^2_B}}},
	\\
	C^{\pm\pm}_{n,m}(k_i) &:= \left| \frac{\hbar v_F\sqrt{g^2(k_i)+\frac{2n}{l^2_B}}\sqrt{g^2(k_i)+\frac{2m}{l^2_B}}}{g(k_i)g'(k_i)\left(E_{n,\pm}(k_i)-E_{m,\pm}(k_i)\right)}\right|,
	\end{align}
Let us simplify this expression in the limit where $T\to 0$. In this limit, the $A_n$s become
\begin{equation}
    A_{n}^{\lambda \lambda'}(k_i)\xrightarrow[]{T\to 0} \theta\left[\mu-E_{n,\lambda}(k_z)\right]-\theta \left[\mu-E_{n+1,\lambda'}(k_i)\right],
\end{equation}
which yield the following
\begin{align}
    A^{++}_n(k_i)&=\theta\left\{ \mu- \hbar v_F \left[h(k_i)+\sqrt{\left(\frac{\omega}{2v_F}-\frac{v_F}{\omega l_B^2}\right)^2}\right]\right\}-\theta\left\{ \mu- \hbar v_F \left[h(k_i)+\sqrt{\left(\frac{\omega}{2v_F}+\frac{v_F}{\omega l_B^2}\right)^2}\right]\right\},
    \\
     A^{--}_n(k_i)&=\theta\left\{ \mu- \hbar v_F \left[h(k_i)-\sqrt{\left(\frac{\omega}{2v_F}-\frac{v_F}{\omega l_B^2}\right)^2}\right]\right\}-\theta\left\{ \mu- \hbar v_F \left[h(k_i)-\sqrt{\left(\frac{\omega}{2v_F}+\frac{v_F}{\omega l_B^2}\right)^2}\right]\right\},
     \\
      A^{+-}_n(k_i)&=\theta\left\{ \mu- \hbar v_F \left[h(k_i)+\sqrt{\left(\frac{\omega}{2v_F}-\frac{v_F}{\omega l_B^2}\right)^2}\right]\right\}-\theta\left\{ \mu- \hbar v_F \left[h(k_i)-\sqrt{\left(\frac{\omega}{2v_F}+\frac{v_F}{\omega l_B^2}\right)^2}\right]\right\},
      \\
       A^{-+}_n(k_i)&=\theta\left\{ \mu- \hbar v_F \left[h(k_i)-\sqrt{\left(\frac{\omega}{2v_F}-\frac{v_F}{\omega l_B^2}\right)^2}\right]\right\}-\theta\left\{ \mu- \hbar v_F \left[h(k_i)+\sqrt{\left(\frac{\omega}{2v_F}+\frac{v_F}{\omega l_B^2}\right)^2}\right]\right\}.
\end{align}
This means that these factors evaluate to
\begin{align}
    A_{n}^{++}(k_i) &= \begin{cases} &1, \quad \hbar v_F \left|\frac{v_F}{\omega l_B^2}-\frac{\omega}{2 v_F}\right| < \mu-\hbar v_F h(k_i)<\hbar v_F \left|\frac{v_F}{\omega l_B^2}+\frac{\omega}{2 v_F}\right| \\ &0, \quad \text{otherwise} \end{cases}
    \\
    A_{n}^{--}(k_i) &= \begin{cases} &-1, \quad -\hbar v_F \left|\frac{v_F}{\omega l_B^2}+\frac{\omega}{2 v_F}\right| < \mu-\hbar v_F h(k_i)<-\hbar v_F \left|\frac{v_F}{\omega l_B^2}-\frac{\omega}{2 v_F}\right| \\ &0, \quad \text{otherwise} \end{cases}
    \\
    A_{n}^{+-}(k_i) &= \begin{cases} &-1, \quad \hbar v_F \left|\frac{v_F}{\omega l_B^2}-\frac{\omega}{2 v_F}\right| < \mu-\hbar v_F h(k_i)<-\hbar v_F \left|\frac{v_F}{\omega l_B^2}+\frac{\omega}{2 v_F}\right| \\ &0, \quad \text{otherwise} \end{cases}
    \\
    A_{n}^{-+}(k_i) &= \begin{cases} &1, \quad -\hbar v_F \left|\frac{v_F}{\omega l_B^2}-\frac{\omega}{2 v_F}\right| < \mu-\hbar v_F h(k_i)<\hbar v_F \left|\frac{v_F}{\omega l_B^2}+\frac{\omega}{2 v_F}\right| \\ &0, \quad \text{otherwise} \end{cases}
\end{align}
Also, the following simplifications can be done to the remaining factors of the response functions,
\begin{align}
    &\quad B_n^+(k_i)B_{n+1}^-(k_i)C_{n,n+1}(k_i) \theta\left(\frac{2 v_F^2}{l_B^2}-\omega^2\right) \nonumber
    \\
    &=\left\{1-\frac{g^2(k_i)}{\frac{v_F^2}{\omega^2l_B^4}-\frac{\omega^2}{4v_F^2}}+g(k_i) \left(\frac{1}{\left|\frac{v_F}{\omega l_B^2}-\frac{\omega}{2v_F}\right|}-\frac{1}{\left| \frac{v_F}{\omega l_B^2}+\frac{\omega}{2v_F}\right|}\right)\right\} \left|\frac{\frac{\omega^2}{4v_F^2}-\frac{v_F^2}{\omega^2l_B^4}}{g(k_i)g'(k_i)\frac{\omega}{v_F}}\right|\theta\left(\frac{2 v_F^2}{l_B^2}-\omega^2\right) \nonumber
    \\
    &= \frac{\frac{2n+1}{l_B^2}-\frac{\omega^2}{2v_F^2}+\tilde{g}(k_i)}{\frac{v_F^2}{\omega^2l_B^4}-\frac{\omega^2}{4v_F^2}} \left|\frac{\frac{\omega^2}{4v_F^2}-\frac{v_F^2}{\omega^2l_B^4}}{g(k_i)g'(k_i)\frac{\omega}{v_F}}\right|\theta\left(\frac{2 v_F^2}{l_B^2}-\omega^2\right) \nonumber
    \\
    &=\frac{\frac{2n+1}{l^2_B}-\frac{\omega^2}{2v_F^2}+\tilde{g}(k_i)}{\left|\tilde{g}(k_i)g'(k_i)\right|}\theta\left(\frac{2 v_F^2}{l_B^2}-\omega^2\right),
    \\
    &\quad B_n^-(k_i)B_{n+1}^+(k_i)C_{n,n+1}(k_i) \theta\left(\frac{2 v_F^2}{l_B^2}-\omega^2\right) \nonumber
    \\
    &=\left\{1-\frac{g^2(k_i)}{\frac{v_F^2}{\omega^2l_B^4}-\frac{\omega^2}{4v_F^2}}+g(k_i) \left(\frac{1}{\left|\frac{v_F}{\omega l_B^2}+\frac{\omega}{2v_F}\right|}-\frac{1}{\left| \frac{v_F}{\omega l_B^2}-\frac{\omega}{2v_F}\right|}\right)\right\} \left|\frac{\frac{\omega^2}{4v_F^2}-\frac{v_F^2}{\omega^2l_B^4}}{g(k_i)g'(k_i)\frac{\omega}{v_F}}\right|\theta\left(\frac{2 v_F^2}{l_B^2}-\omega^2\right) \nonumber
    \\
    &= \frac{\frac{2n+1}{l_B^2}-\frac{\omega^2}{2v_F^2}-\tilde{g}(k_i)}{\frac{v_F^2}{\omega^2l_B^4}-\frac{\omega^2}{4v_F^2}} \left|\frac{\frac{\omega^2}{4v_F^2}-\frac{v_F^2}{\omega^2l_B^4}}{g(k_i)g'(k_i)\frac{\omega}{v_F}}\right|\theta\left(\frac{2 v_F^2}{l_B^2}-\omega^2\right) \nonumber
    \\
    &=\frac{\frac{2n+1}{l^2_B}-\frac{\omega^2}{2v_F^2}-\tilde{g}(k_i)}{\left|\tilde{g}(k_i)g'(k_i)\right|}\theta\left(\frac{2 v_F^2}{l_B^2}-\omega^2\right),
     \\
    &\quad B_n^+(k_i)B_{n+1}^+(k_i)C_{n,n+1}(k_i) \theta\left(\omega^2-\frac{2 v_F^2}{l_B^2}\right) \nonumber
    \\
    &=\left\{1+\frac{g^2(k_i)}{\frac{\omega^2}{4v_F^2}-\frac{v_F^2}{\omega^2l_B^4}}+g(k_i) \left(\frac{1}{\left|\frac{v_F}{\omega l_B^2}+\frac{\omega}{2v_F}\right|}+\frac{1}{\left| \frac{v_F}{\omega l_B^2}-\frac{\omega}{2v_F}\right|}\right)\right\} \left|\frac{\frac{\omega^2}{4v_F^2}-\frac{v_F^2}{\omega^2l_B^4}}{g(k_i)g'(k_i)\frac{\omega}{v_F}}\right|\theta\left(\omega^2-\frac{2 v_F^2}{l_B^2}\right) \nonumber
    \\
    &= \frac{\frac{\omega^2}{2v_F^2}-\frac{2n+1}{l_B^2}+\tilde{g}(k_i)}{\frac{\omega^2}{4v_F^2}-\frac{v_F^2}{\omega^2l_B^4}} \left|\frac{\frac{\omega^2}{4v_F^2}-\frac{v_F^2}{\omega^2l_B^4}}{g(k_i)g'(k_i)\frac{\omega}{v_F}}\right|\theta\left(\omega^2-\frac{2 v_F^2}{l_B^2}\right) \nonumber
    \\
    &=\frac{\frac{\omega^2}{2v_F^2}-\frac{2n+1}{l^2_B}+\tilde{g}(k_i)}{\left|\tilde{g}(k_i)g'(k_i)\right|}\theta\left(\omega^2-\frac{2 v_F^2}{l_B^2}\right),
    \\
    &\quad B_n^-(k_i)B_{n+1}^-(k_i)C_{n,n+1}(k_i) \theta\left(\omega^2-\frac{2 v_F^2}{l_B^2}\right) \nonumber
    \\
    &=\left\{1+\frac{g^2(k_i)}{\frac{\omega^2}{4v_F^2}-\frac{v_F^2}{\omega^2l_B^4}}-g(k_i) \left(\frac{1}{\left|\frac{v_F}{\omega l_B^2}+\frac{\omega}{2v_F}\right|}+\frac{1}{\left| \frac{v_F}{\omega l_B^2}-\frac{\omega}{2v_F}\right|}\right)\right\} \left|\frac{\frac{\omega^2}{4v_F^2}-\frac{v_F^2}{\omega^2l_B^4}}{g(k_i)g'(k_i)\frac{\omega}{v_F}}\right|\theta\left(\omega^2-\frac{2 v_F^2}{l_B^2}\right) \nonumber
    \\
    &= \frac{\frac{\omega^2}{2v_F^2}-\frac{2n+1}{l_B^2}-\tilde{g}(k_i)}{\frac{\omega^2}{4v_F^2}-\frac{v_F^2}{\omega^2l_B^4}} \left|\frac{\frac{\omega^2}{4v_F^2}-\frac{v_F^2}{\omega^2l_B^4}}{g(k_i)g'(k_i)\frac{\omega}{v_F}}\right|\theta\left(\omega^2-\frac{2 v_F^2}{l_B^2}\right) \nonumber
    \\
    &=\frac{\frac{\omega^2}{2v_F^2}-\frac{2n+1}{l^2_B}-\tilde{g}(k_i)}{\left|\tilde{g}(k_i)g'(k_i)\right|}\theta\left(\omega^2-\frac{2 v_F^2}{l_B^2}\right),
\end{align}
with $\tilde{g}(k_i):=\left|\frac{\omega}{v_F}\right|g(k_i)$. Then, we arrive at,
\begin{align}
    &\text{Im}\left[\chi_{xx}(\omega)\right]=-\frac{e^2v_F^2}{16\pi l^2_B}\sum_{n=0}^{n_{\text{max}}}\sum_{i=1}^{2m} \nonumber
    \\
    &\left\{ A_n^{++}(k_i)\left[\frac{\frac{2n+1}{l^2_B}-\frac{\omega^2}{2v_F^2}+\tilde{g}(k_i)}{\left|\tilde{g}(k_i)g'(k_i)\right|}\right]\theta\left(\frac{2 v_F^2}{l_B^2}-\omega^2\right)\sgn{(-\omega)}\right.\nonumber
    \\
    &+A^{--}_n(k_i)\left[\frac{\frac{2n+1}{l^2_B}-\frac{\omega^2}{2v_F^2}-\tilde{g}(k_i)}{\left|\tilde{g}(k_i)g'(k_i)\right|}\right]\theta\left(\frac{2 v_F^2}{l_B^2}-\omega^2\right)\sgn{(\omega)}\nonumber
    \\
    & +A_n^{+-}(k_i)\left[\frac{\frac{\omega^2}{2v_F^2}-\frac{2n+1}{l^2_B}+\tilde{g}(k_i)}{\left|\tilde{g}(k_i)g'(k_i)\right|}\right]\theta\left(\omega^2-\frac{2 v_F^2}{l_B^2}\right)\sgn{(\omega)}\nonumber
    \\
    &\left. +A^{-+}_n(k_i)\left[\frac{\frac{\omega^2}{2v_F^2}-\frac{2n+1}{l^2_B}-\tilde{g}(k_i)}{\left|\tilde{g}(k_i)g'(k_i)\right|}\right]\theta\left(\omega^2-\frac{2 v_F^2}{l_B^2}\right)\sgn{(-\omega)} \right\},
    \end{align}
    \begin{align}
    &\text{Re}\left[\chi_{xy}(\omega)\right]=-\frac{e^2v_F^2}{16\pi l^2_B}\sum_{n=0}^{n_{\text{max}}}\sum_{i=1}^{2m} \nonumber
    \\
    &\left\{ A_n^{++}(k_i)\left[\frac{\frac{2n+1}{l^2_B}-\frac{\omega^2}{2v_F^2}+\tilde{g}(k_i)}{\left|\tilde{g}(k_i)g'(k_i)\right|}\right]\theta\left(\frac{2 v_F^2}{l_B^2}-\omega^2\right)\right.\nonumber
    \\
    &+A^{--}_n(k_i)\left[\frac{\frac{2n+1}{l^2_B}-\frac{\omega^2}{2v_F^2}-\tilde{g}(k_i)}{\left|\tilde{g}(k_i)g'(k_i)\right|}\right]\theta\left(\frac{2 v_F^2}{l_B^2}-\omega^2\right) \nonumber
    \\
    & +A_n^{+-}(k_i)\left[\frac{\frac{\omega^2}{2v_F^2}-\frac{2n+1}{l^2_B}+\tilde{g}(k_i)}{\left|\tilde{g}(k_i)g'(k_i)\right|}\right]\theta\left(\omega^2-\frac{2 v_F^2}{l_B^2}\right)\nonumber
    \\
    &\left. +A^{-+}_n(k_i)\left[\frac{\frac{\omega^2}{2v_F^2}-\frac{2n+1}{l^2_B}-\tilde{g}(k_i)}{\left|\tilde{g}(k_i)g'(k_i)\right|}\right]\theta\left(\omega^2-\frac{2 v_F^2}{l_B^2}\right)\right\}.
\end{align}

\subsection{Contribution from the Chiral Level}
Even though the contribution from the chiral level is, at least schematically, covered by the reasoning above, it serves a purpose to spell it out explicitly. Initially, this reads,
\begin{align}
    &\quad \text{Im}\left[\chi_{xx}^0(\omega)\right] =-\frac{e^2v_F^2}{8 \pi l_B^2}\int dk_z\nonumber
    \\
    &\left(\left\{\theta\left[\mu-E_0(k_z)\right]-\theta\left[\mu-E_{1,+}(k_z)\right]\right\}\left[1-\frac{g(k_z)}{\sqrt{g^2(k_z)+\frac{2}{l_B^2}}}\right] \delta\left\{\hbar\omega-\hbar v_F\left[g(k_z)-\sqrt{g^2(k_z)+\frac{2}{l_B^2}}\right]\right\} \right. \nonumber
    \\
    &+\left\{\theta\left[\mu-E_0(k_z)\right]-\theta\left[\mu-E_{1,-}(k_z)\right]\right\}\left[1+\frac{g(k_z)}{\sqrt{g^2(k_z)+\frac{2}{l_B^2}}}\right]\delta\left\{\hbar\omega-\hbar v_F\left[g(k_z)+\sqrt{g^2(k_z)+\frac{2}{l_B^2}}\right]\right\} \nonumber
    \\
    &+\left\{\theta\left[\mu-E_{1,+}(k_z)\right]-\theta\left[\mu-E_0(k_z)\right]\right\}\left[1-\frac{g(k_z)}{\sqrt{g^2(k_z)+\frac{2}{l_B^2}}}\right]\delta\left\{\hbar\omega-\hbar v_F\left[\sqrt{g^2(k_z)+\frac{2}{l_B^2}}-g(k_z)\right]\right\} \nonumber
    \\
    &+\left.\left\{\theta\left[\mu-E_{1,-}(k_z)\right]-\theta\left[\mu-E_0(k_z)\right]\right\}\left[1+\frac{g(k_z)}{\sqrt{g^2(k_z)+\frac{2}{l_B^2}}}\right]\delta\left\{\hbar\omega-\hbar v_F\left[-\sqrt{g^2(k_z)+\frac{2}{l_B^2}}-g(k_z)\right]\right\}\right),
    \\
     &\quad \text{Re}\left[\chi_{xy}^0(\omega)\right] =-\frac{e^2v_F^2}{8 \pi l_B^2}\int dk_z\nonumber
    \\
    &\left(\left\{\theta\left[\mu-E_0(k_z)\right]-\theta\left[\mu-E_{1,+}(k_z)\right]\right\}\left[1-\frac{g(k_z)}{\sqrt{g^2(k_z)+\frac{2}{l_B^2}}}\right] \delta\left\{\hbar\omega-\hbar v_F\left[g(k_z)-\sqrt{g^2(k_z)+\frac{2}{l_B^2}}\right]\right\} \right. \nonumber
    \\
    &+\left\{\theta\left[\mu-E_0(k_z)\right]-\theta\left[\mu-E_{1,-}(k_z)\right]\right\}\left[1+\frac{g(k_z)}{\sqrt{g^2(k_z)+\frac{2}{l_B^2}}}\right]\delta\left\{\hbar\omega-\hbar v_F\left[g(k_z)+\sqrt{g^2(k_z)+\frac{2}{l_B^2}}\right]\right\} \nonumber
    \\
    &-\left\{\theta\left[\mu-E_{1,+}(k_z)\right]-\theta\left[\mu-E_0(k_z)\right]\right\}\left[1-\frac{g(k_z)}{\sqrt{g^2(k_z)+\frac{2}{l_B^2}}}\right]\delta\left\{\hbar\omega-\hbar v_F\left[\sqrt{g^2(k_z)+\frac{2}{l_B^2}}-g(k_z)\right]\right\} \nonumber
    \\
    &-\left.\left\{\theta\left[\mu-E_{1,-}(k_z)\right]-\theta\left[\mu-E_0(k_z)\right]\right\}\left[1+\frac{g(k_z)}{\sqrt{g^2(k_z)+\frac{2}{l_B^2}}}\right]\delta\left\{\hbar\omega-\hbar v_F\left[-\sqrt{g^2(k_z)+\frac{2}{l_B^2}}-g(k_z)\right]\right\}\right).
\end{align}
The integrals are again carried out by solving the expressions inside the Dirac $\delta$-distributions, which yields,
\begin{align}
    \hbar\omega-\hbar v_F\left[g(k_z)-\sqrt{g^2(k_z)+\frac{2}{l_B^2}}\right]&=0 \Rightarrow g(k_z) = \frac{\omega^2l_B^2-2v_F^2}{2l_B^2v_F\omega}\theta\left(-\omega\right),\label{eq:k01}
    \\
    \hbar\omega-\hbar v_F\left[g(k_z)+\sqrt{g^2(k_z)+\frac{2}{l_B^2}}\right] &=0 \Rightarrow g(k_z)=\frac{\omega^2l_B^2-2v_F^2}{2l_B^2v_F\omega}\theta\left(\omega\right),\label{eq:k02}
    \\
    \hbar\omega-\hbar v_F\left[\sqrt{g^2(k_z)+\frac{2}{l_B^2}}-g(k_z)\right] &=0\Rightarrow g(k_z)=\frac{2v_F^2-\omega^2l_B^2}{2l_B^2v_F\omega}\theta\left(\omega\right),\label{eq:k03}
    \\
    \hbar\omega-\hbar v_F\left[-\sqrt{g^2(k_z)+\frac{2}{l_B^2}}-g(k_z)\right] &=0\Rightarrow g(k_z)=\frac{2v_F^2-\omega^2l_B^2}{2l_B^2v_F\omega}\theta\left(-\omega\right). \label{eq:k04}
\end{align}
Plugging these solution back into the expression for the chiral contribution, we arrive at
\begin{align}
     &\quad \text{Im}\left[\chi_{xx}^0(\omega)\right] = \nonumber
     \\
     &-\frac{e^2v_F^2}{8 \pi l_B^2}\sum_{i=1}^{m}\left(\left\{\theta\left[\mu-E_0(k^i_1)\right]-\theta\left[\mu-E_{1,+}(k^i_1)\right]\right\}\frac{\theta(-\omega)}{\hbar v_F |g'(k^i_1)|}+\left\{\theta\left[\mu-E_0(k^i_1)\right]-\theta\left[\mu-E_{1,-}(k^i_1)\right]\right\}\frac{\theta(\omega)}{\hbar v_F |g'(k^i_1)|} \right. \nonumber
     \\
     &\quad \left.\left\{\theta\left[\mu-E_{1,+}(k^i_2)\right]-\theta\left[\mu-E_0(k^i_2)\right]\right\}\frac{\theta(\omega)}{\hbar v_F |g'(k^i_2)|} + \left\{\theta\left[\mu-E_{1,-}(k^i_2)\right]-\theta\left[\mu-E_0(k^i_2)\right]\right\}\frac{\theta(-\omega)}{\hbar v_F |g'(k^i_2)|}\right),
     \\
     &\quad \text{Re}\left[\chi_{xy}^0(\omega)\right] = \nonumber
     \\
     &-\frac{e^2v_F^2}{8 \pi l_B^2}\sum_{i=1}^{m}\left(\left\{\theta\left[\mu-E_0(k^i_1)\right]-\theta\left[\mu-E_{1,+}(k^i_1)\right]\right\}\frac{1}{\hbar v_F |g'(k^i_1)|}+\left\{\theta\left[\mu-E_0(k^i_1)\right]-\theta\left[\mu-E_{1,-}(k^i_1)\right]\right\}\frac{1}{\hbar v_F |g'(k^i_1)|} \right. \nonumber
     \\
     &\quad \left.\left\{\theta\left[\mu-E_{1,+}(k^i_2)\right]-\theta\left[\mu-E_0(k^i_2)\right]\right\}\frac{1}{\hbar v_F |g'(k^i_2)|} + \left\{\theta\left[\mu-E_{1,-}(k^i_2)\right]-\theta\left[\mu-E_0(k^i_2)\right]\right\}\frac{1}{\hbar v_F |g'(k^i_2)|}\right),
\end{align}
where $k_{1,2}^i$ denotes the solutions obtained when solving Eqs.~\eqref{eq:k01}-\eqref{eq:k04} for $k_z$.

\end{widetext}


\begin{thebibliography}{9}
\bibitem{Klitzing1980}
K. von Klitzing, \href{https://doi.org/10.1103/PhysRevLett.45.494}{Phys. Rev. Lett. {\bf 45}, 494 (1980)}.

\bibitem{Altland1997}
A. Altland, and M.R. Zirnbauer, \href{https://doi.org/10.1103/PhysRevB.55.1142}{Phys. Rev. B {\bf 55}, 1142 (1997)}.

\bibitem{hasankane}
M.Z. Hasan and C.L. Kane, \href{https://journals.aps.org/rmp/abstract/10.1103/RevModPhys.82.3045}{Rev. Mod. Phys. {\bf 82}, 3045 (2010)}.

\bibitem{qizhang}
X.-L. Qi and S.-C. Zhang, \href{https://journals.aps.org/rmp/abstract/10.1103/RevModPhys.83.1057}{Rev. Mod. Phys. {\bf 83}, 1057 (2011)}.

\bibitem{goerbig}
M.O. Goerbig, \href{http://dx.doi.org/10.1103/RevModPhys.83.1193}{Rev. Mod. Phys. {\bf 83}, 1193 (2011)}.

\bibitem{weylreview}
N.P. Armitage, E.J. Mele, and A. Vishwanath, \href{https://journals.aps.org/rmp/abstract/10.1103/RevModPhys.90.015001}{Rev. Mod. Phys. {\bf 90}, 015001 (2018)}.

\bibitem{XBAN2015} 
S.-Y. Xu, I. Belopolski, N. Alidoust, M. Neupane, G. Bian, C. Zhang, R. Sankar, G. Chang, Z. Yuan, C.-C. Lee, S.-M. Huang, H. Zheng, J. Ma, D. S. Sanchez, B. Wang, A. Bansil, F. Chou, P. P. Shibayev, H. Lin, S. Jia, and M. Z. Hasan, \href{https://science.sciencemag.org/content/349/6248/613.abstract}{Science {\bf 349}, 613 (2015)}.

\bibitem{LWFM2015}
B.Q. Lv, H.M. Weng, B.B. Fu, X.P. Wang, H. Miao, J. Ma, P. Richard, X.C. Huang, L.X. Zhao, G.F. Chen, Z. Fang, X. Dai, T. Qian, and H. Ding, \href{https://journals.aps.org/prx/abstract/10.1103/PhysRevX.5.031013}{Phys. Rev. X {\bf 5}, 031013 (2015)}.

\bibitem{LWYRFJS2015}
L. Lu, Z Wang, D. Ye, L. Ran, L. Fu, J.D. Joannopoulos, and M. Solja\v{c}i\'c, \href{http://dx.doi.org/10.1126/science.aaa9273}{Science {\bf 349}, 6248, 622-624 (2015)}.

\bibitem{W1929} H. Weyl, \href{https://link.springer.com/article/10.1007/BF01339504}{Z. Physik {\bf 56}, 330 (1929)}.

\bibitem{HZLW2015}
X. Huang, L. Zhao, Y. Long, P. Wang, D. Chen, Z. Yang, H. Lang, M. Xue, H. Weng, Z. Fang, X. Dai, and G. Chen, \href{http://dx.doi.org/10.1103/physrevx.5.031023}{Phys. Rev. X {\bf 5}, 031023 (2015)}.

\bibitem{WTVS2011} X. Wan, A.M. Turner, A. Vishwanath, and S.Y. Savrasov, \href{http://link.aps.org/doi/10.1103/PhysRevB.83.205101}{Phys. Rev. B {\bf 83}, 205101 (2011)}.

\bibitem{HQ2013} P. Hosur, and X. Qi \href{https://doi.org/10.1016/j.crhy.2013.10.010}{Comptes Rendus Physique {\bf 14}, 9-10, (2013)}.

\bibitem{SGWWTDB2015} 
A.A. Soluyanov, D. Gresch, Z. Wang, Q.S. Wu, M. Troyer, X. Dai, and B.A. Bernevig, \href{http://dx.doi.org/10.1038/nature15768}{Nature {\bf 527}, 495-498 (2015)}.

\bibitem{BLTMU2015} 
E.J. Bergholtz, Z. Liu, M. Trescher, R. Moessner, and M. Udagawa, \href{http://link.aps.org/doi/10.1103/PhysRevLett.114.016806}{Phys. Rev. Lett {\bf 114}, 016806 (2015)}.

\bibitem{AXFT2014} 
M.N. Ali, J. Xiong, S. Flynn, J. Tao, Q.D. Gibson, L.M. Schoop, T. Liang, N. Haldolaarachchige, M. Hirschenberger, N.P. Ong, and R.J. Cava, \href{https://www.nature.com/articles/nature13763}{Nature {\bf 514}, 205-208 (2014)}.

\bibitem{KSXMY2017} 
N. Kumar, Y. Sun, N. Xu, K. Manna, M. Yao, V.S\"uss, I. Leermak, O. Young, T.F\"orster, M. Schmidt, H. Borrmann, B. Yan, U. Zeitler, M. Shi, C. Felser, and C. Shekhar, \href{https://doi.org/10.1038/s41467-017-01758-z}{Nature Comms {\bf 8} 1642 (2017)}.

\bibitem{Zetal2016} 
L.-K. Zeng, R. Lou, D.-S. Wu, Q.N. Xu, P.-J. Guo, L.-Y. Kong, Y.-G. Zhong, J.-Z. Ma, B.-B. Fu, P. Richard, P. Wang, G.T. Liu, L. Lu, Y.-B. Huang, C. Fang, S.-S. Sun, Q. Wang, L. Wang, Y.-G. Shi, H.M. Weng, H.-C. Lei, K. Liu, S.-C. Wang, T. Qian, J.-L. Luo, and H. Ding, \href{https://doi.org/10.1103/physrevlett.117.127204}{Phys. Rev. Lett. {\bf 117}, 127204 (2016)}.

\bibitem{ZXBY2016}
C.-L. Zhang, S.-Y. Xu, I. Belopolski, Z. Yuan, Z. Lin, B. Tong, G. Bian, N. Alidoust, C.-C. Lee, S.-M. Huang, T.-R. Chang, G. Chang, C.-H. Hsu, H.-R. Jeng, M. Neupance, D.S. Sanchez, H. Zheng, J. Wang, H. Lin, C. Zhang, H.-Z. Lu, S.-Q. Shen, T. Neupert, M.Z. Hasan, and S. Jia, \href{http://dx.doi.org/10.1038/ncomms10735}{Nature Comms. {\bf 7} 10735 (2016)}.

\bibitem{NM1983}
H.B. Nielsen, and M. Ninomiya, \href{https://doi.org/10.1016/0370-2693(83)91529-0}{Phys. Lett. B {\bf 130}, 6, 389-396 (1983)}.

\bibitem{XPYQ2015}
C.-X. Liu, P. Ye, and X.-L. Qi, \href{https://journals.aps.org/prb/abstract/10.1103/PhysRevB.92.119904}{Phys. Rev. B {\bf 92}, 119904 (2015)}.

\bibitem{SY2012}
D.T. Son, and N .Yamamoto, \href{https://link.aps.org/doi/10.1103/PhysRevLett.109.181602}{Phys. Rev. Lett. {\bf 109}, 181602 (2012)}.

\bibitem{G2012} 
A.G. Grushin, \href{http://dx.doi.org/10.1103/PhysRevD.86.045001}{Phys. Rev. D {\bf 86}, 045001 (2012)}.

\bibitem{ZB2012}
A.A. Zyuzin and A.A. Burkov, \href{https://link.aps.org/doi/10.1103/PhysRevB.86.115133}{Phys. Rev. B {\bf 86}, 115133 (2012)}.

\bibitem{GT2013}
P. Goswami and S. Tewari, \href{https://link.aps.org/doi/10.1103/PhysRevB.88.245107}{Phys. Rev. B {\bf 88}, 245107 (2013)}.

\bibitem{PGAPV2014} 
S. Parameswaran, T. Grover, D.A. Abanin, D.A. Pesin, and A. Vishwanath, \href{https://doi.org/10.1103/physrevx.4.031035}{Phys. Rev. X {\bf 4}, 031035 (2014)}.

\bibitem{KGM2015}
J. Klier, I.V. Gornyi, and A.D. Mirlin, \href{https://doi.org/10.1103/physrevb.92.205113}{Phys. Rev. B {\bf 92}, 205113 (2015)}.

\bibitem{KGM2017}
J. Klier, I.V. Gornyi, and A.D. Mirlin, \href{https://link.aps.org/doi/10.1103/PhysRevB.96.214209}{Phys. Rev. B {\bf 96}, 214209 (2017)}.

\bibitem{BKS2018}
J. Behrends, F.K. Kunst, and B. Sbierski, \href{https://link.aps.org/doi/10.1103/PhysRevB.97.064203}{Phys. Rev. B {\bf 97}, 064203 (2018)}.

\bibitem{Huangetal2015}
X. Huang, L. Zhao, Y. Long, P. Wang, D. Chen, Z. Yang, H. Liang, M. Xue, H. Weng, Z. Fang, X. Dai, and G. Chen, \href{http://dx.doi.org/10.1103/physrevx.5.031023}{Phys. Rev. X {\bf 5}, 031023 (2015)}.

\bibitem{AC2013}
P.E.C. Ashby, J.P. Carbotte, \href{https://journals.aps.org/prb/abstract/10.1103/PhysRevB.87.245131}{Phys. Rev. B {\bf 87}, 245131 (2013)}.

\bibitem{Tchoumakov2016}
S. Tchoumakov, M. Civelli, and M.O. Goerbig, \href{https://doi.org/10.1103/PhysRevLett.117.086402}{Phys. Rev. Lett. {\bf 117}, 086402 (2016)}.

\bibitem{Stalhammar20}
M. St{\aa}lhammar, J. Larana-Aragon, J. Knolle, and E.J. Bergholtz, \href{}{https://doi.org/10.1103/PhysRevB.102.235134}.

\bibitem{Yadav2022}
S. Yadav, S. Sekh, I. Mandal, \href{https://doi.org/10.1016/j.physb.2023.414765}{Physica B: Condensed Matter {\bf 656}, 414765 (2023)}.

\bibitem{XZC2018} 
X. Yuan, Z. Yan, C. Song, M. Zhang, Z. Li, C. Zhang, Y. Liu, W. Wang, M. Zhao, Z. Lin, T. Xie, J. Ludwig, Y. Jiang, X. Zhang, C. Shang, Z. Ye, J. Wang, F. Chen, Z. Xia, D. Smirnov, X. Chen, Z. Wang, H. Yan, and F. Xiu, \href{https://www.nature.com/articles/s41467-018-04080-4}{Nature Comm. {\bf 9}, 1854 (2018)}.

\bibitem{PD2020}
A.V. Pronin, and M. Dressel, \href{https://onlinelibrary.wiley.com/doi/full/10.1002/pssb.202000027}{Physica Status Solidi B 00027 (2020)}.

\bibitem{PGW2020}
S. Polatkan, M.O. Goerbig, J. Wyzula, R. Kemmler, L.Z. Maulana, B.A. Piot, I. Crassee, A. Akrap, C. Shekhar, C. Felser, M. Dressel, A.V. Pronin, and M. Orlita, \href{https://journals.aps.org/prl/abstract/10.1103/PhysRevLett.124.176402}{Phys. Rev. Lett. {\bf 124}, 176402 (2020)}.

\bibitem{UB2016}
M. Udagawa, and E.J. Bergholtz, \href{https://journals.aps.org/prl/abstract/10.1103/PhysRevLett.117.086401}{Phys. Rev. Lett. {\bf 117}, 086401 (2016)}.

\bibitem{YYY2016}Z.-M. Yu, Y. Yao, and S.A. Yang, \href{https://journals.aps.org/prl/abstract/10.1103/PhysRevLett.117.077202}{Phys. Rev. Lett. {\bf 117}, 077202 (2016)}.

\bibitem{TBUK2017}M. Trescher, E.J. Bergholtz, M. Udagawa, and J. Knolle, \href{https://journals.aps.org/prb/abstract/10.1103/PhysRevB.96.201101}{Phys. Rev. B {\bf 96}, 201101(R) (2017)}.

\bibitem{TBK2018}
M. Trescher, E.J. Bergholtz, and J. Knolle, \href{https://journals.aps.org/prb/abstract/10.1103/PhysRevB.98.125304}{Phys. Rev. B {\bf 98}, 125304 (2018)}.

\bibitem{Xiong2022}
F. Xiong, C. Honerkamp, D.M. Kennes, and T. Nag, \href{https://doi.org/10.1103/PhysRevB.106.045424}{Phys. Rev. B {\bf 106}, 045424 (2022)}.

\bibitem{Chang2023}
M. Chang, R. Ma, and L. Sheng, \href{https://doi.org/10.1103/PhysRevB.108.165416}{Phys. Rev. B {\bf 108}, 165416 (2023)}.

\bibitem{Lu2015}
H.-Z. Lu, S.-B. Zhang, and S.-Q. Shen, \href{https://doi.org/10.1103/PhysRevB.92.045203}{Phys. Rev. B {\bf 92}, 045203 (2015)}.

\bibitem{Li2016}
X. Li, B. Roy, and S. Das Sarma, \href{https://doi.org/10.1103/PhysRevB.94.195144}{Phys. Rev. B {\bf 94}, 195144 (2016)}.

\bibitem{Wang2017}
C.M. Wang, H.-P. Sun, H.-Z. Lu, and X.C. Xie, \href{https://doi.org/10.1103/PhysRevLett.119.136806}{Phys. Rev. Lett. {\bf 119}, 136806 (2017)}.

\bibitem{Li2020}
H. Li, H. Liu, H. Jiang, and X.C. Xie, \href{https://doi.org/10.1103/PhysRevLett.125.036602}{Phys. Rev. Lett. {\bf 125}, 036602 (2020)}.







\end{thebibliography}
\end{document}